\documentclass{eethesis}

\usepackage{amssymb}

\usepackage[cmex10]{amsmath}
\usepackage{fixltx2e}
\usepackage{stfloats}
\usepackage{graphicx}

\begin{document}
\pagenumbering{roman}
\maketitlepage{On Randomized Sensing and Access Schemes in \\ Wireless Ad-Hoc Cognitive Networks} {Armin Banaei}{Master of Science}{October 2009}{Electrical and Computer Engineering}

\abs{On Randomized Sensing and Access Schemes in \\ Wireless Ad-Hoc Cognitive Networks}{October 2009}{Armin Banaei}{M.S., Texas A\&M University}{Costas Georghiades}{Over the past decade we have witnessed a rapid growth and development in wireless communication systems, to the point that conventional spectrum allocation policies may not be able to fulfill them all. Federal Communications Commission (FCC) licenses certain frequency segments to a particular user in a particular geographic area. Industrial, Scientific and Medical (ISM) radio bands have also been envisioned for all other unlicensed user to share, as long as they follow certain power regulations. But with the recent boom in the wireless technologies, these open channels have become overcrowded with everything from wireless networks to wireless controllers.

Therefore, the regulatory and standardization agencies have been working on new spectrum regulation policies for wireless communication systems. The underlying idea is to let unlicensed users to use the licensed band as long as they can guarantee low interference to the licensed users. Though seemingly simple, sophisticated interference management protocols are needed to meet the expected level of transparency accepted by licensed users. In this thesis we adopt the dynamic spectrum access approach to limit the interference to primary users and analyze the performance of the cognitive MAC protocols based on randomized sensing and access schemes.

We assume that the secondary users are equipped with spectrum sensors. We consider two main settings; In the first setting we assume that the secondary users deploy the ALOHA protocol to resolve the contentions among each other and can only sense and access a single frequency band at a time. We introduce a heuristic sensing scheme and analyze the performance of the system in two scenarios; In the first one we conduct our analysis over the data-link layer performance of the protocol. In the second one we analyze the physical layer performance of the system, in which we demonstrate the trade-off between the secondary network throughput and the interference level to primary network. Furthermore, we conjecture that the optimal number of secondary users that can co-exist with the primary network is proportional to the \emph{frequency diversity} of the system, the \emph{average backs-off number} of the secondary users and the \emph{spatial diversity} in the secondary network.

In the second setting we assume that secondary users utilize the CSMA/CA protocol to avoid collisions among themselves and analyze the performance of the system only in data-link layer\footnote{The physical layer analysis could follow the same steps as in the first setting.}. We investigate how much benefit we would gain by sensing and accessing multiples at a time channel and determine the associated optimal sensing and access protocols. We conclude the analysis by considering the role of detection errors in the design of the optimal cognitive MAC protocols.}

\pagestyle{headings}
\setlength{\headheight}{36pt}
\tableofcontents
\listoftables
\listoffigures
\clearpage
\pagenumbering{arabic}
\setlength{\headheight}{12pt}
\pagestyle{myheadings}

\chapter{Introduction}
\body

\section{Background}

The demand for wireless services has been growing rapidly throughout the world in the last decade. The extensive use of voice over IP networks, gaming consoles, PDA's and Wi-Fi networks has shown that wide-band wireless communication is becoming more and more popular and demanded as the last mile connection rather than cable, fiber and etc.

\begin{figure}
\centering
\includegraphics[width=6 in]{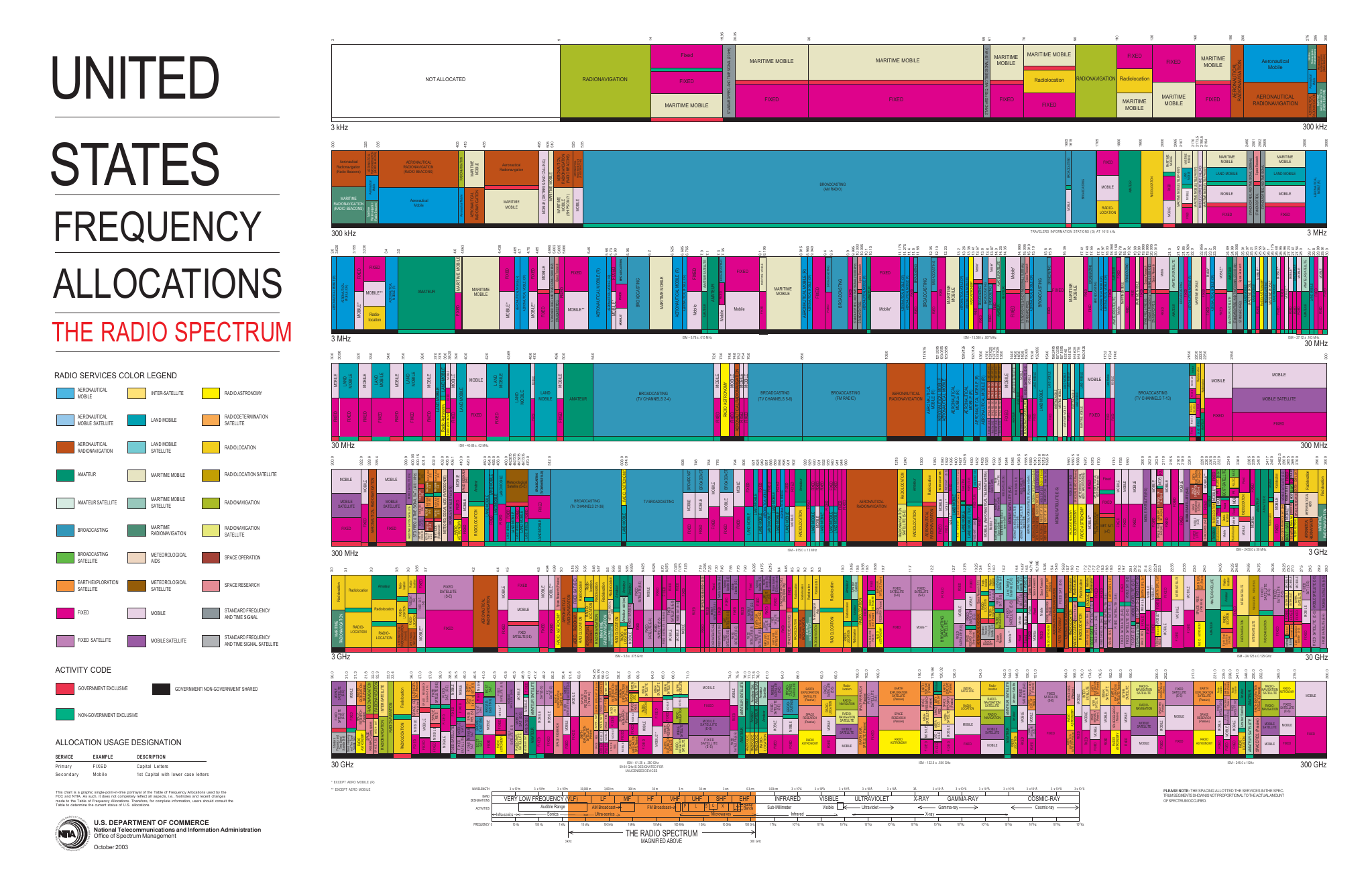}
\caption{NTIA's chart of frequency allocation}
\label{fig:allochrt}
\end{figure}

\begin{figure}
\centering
\includegraphics [width=5in]{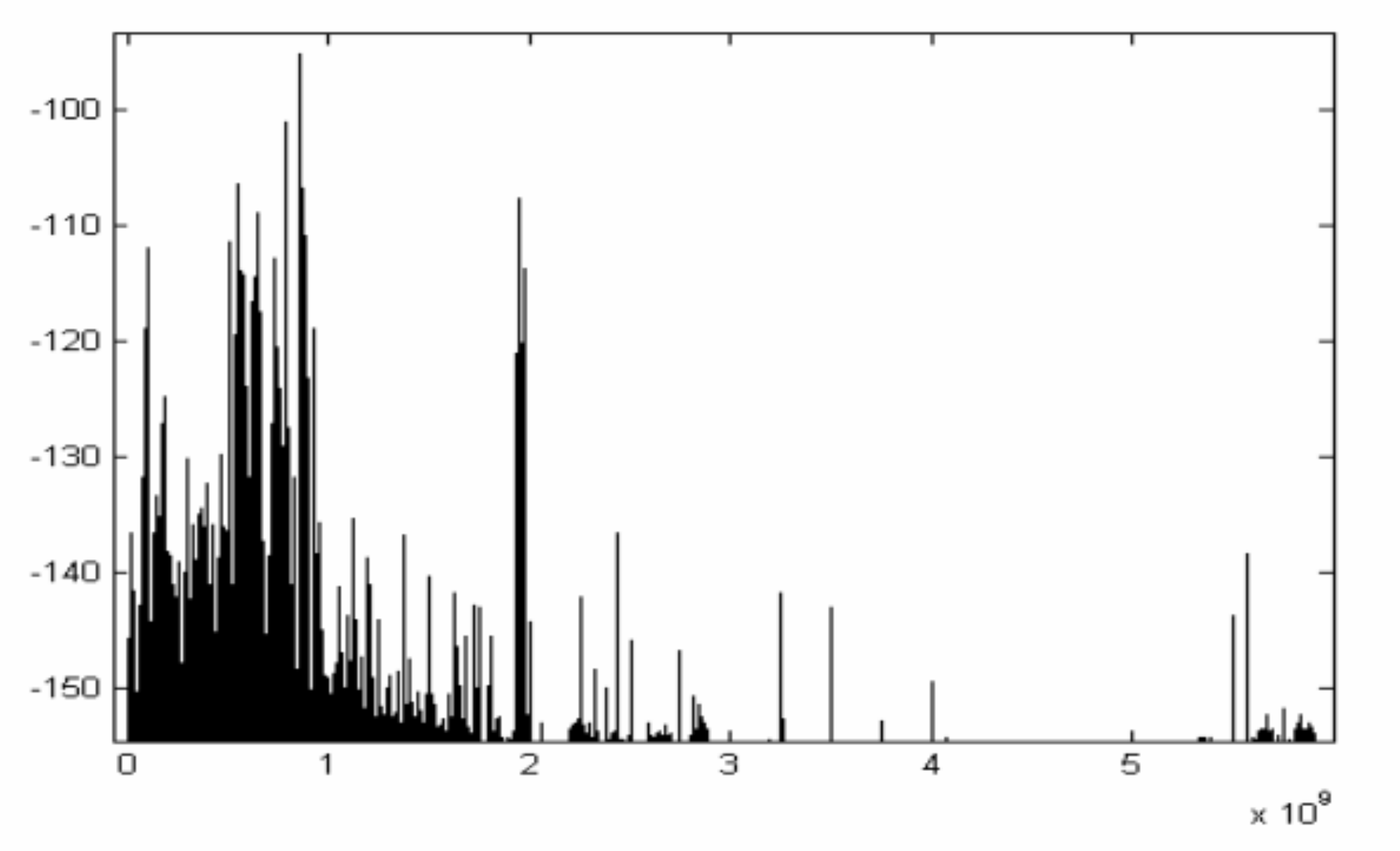}
\caption{Measurement of 0-6 GHz spectrum utilization at Berkeley Wireless Research Center; Power Spectral Density (PSD) of the received 6 GHz wide signal collected for a span of 50ìs sampled at 20 GS/s}
\label{fig:BWRC}
\end{figure}

Conventionally, FCC has allocated each frequency band exclusively to a single user within a geographical region and prohibits the transmission of of other unlicensed users in that band. However, with the emergence of personal wireless communications, static spectrum allocation is no longer reasonable due to economical and technological factors. Therefore, Industrial, Scientific and medical (ISM) bands have been provided to support unlicensed networks at 900 MHz, 2.4 GHz, and 5.8 GHz which has become more and more crowded. Figure \ref{fig:allochrt} shows the NTIA's chart of frequency allocation \cite{NTIA}. From the chart it appears that almost all usable frequency spectrum has been allocated and we are running out of spectrum. However, measurements of actual spectrum usage shows an abundance of  idle frequency bands in the seemingly crowded radio spectrum \cite{speceffic} \cite{Spectrumwhitespacemeasurements}. Figure \ref{fig:BWRC} shows the actual measurements taken in downtown Berkeley, which are believed to be typical and indicates low utilization, especially in the 3-6 MHz bands \cite{Implementationissues}. Another example could be the current television broadcast frequency bands where on the average only $8$ channels out of the 68 allocated channels are being used in any given TV market \cite{Clancy}, which suggests an utilization factor of about 12\%. As such, a change to the current spectrum allocation  policies is desired. However, there is still a strong debate going on among economists about the approaches to fix this problem. Some suggest that the introduction of a secondary market in the already existing market could greatly reduce the inefficiencies in spectrum usage \cite{Spectrummanagement}, while others believe that a common band for all the users is the best solution \cite{Overcomingagoraphobia}. They argue that ``wireless transmissions can be regulated by a combination of (a) baseline rules that allow users to coordinate their use, to avoid interference-producing collisions, and to prevent, for the most part, congestion, by conforming to equipment manufacturer’s specifications, and (b) industry and government sponsored standards'' \cite{Overcomingagoraphobia}. More specifically, in the first proposed approach primary or licensed users have a higher priority and secondary or unlicensed users have a lower priority in accessing the spectrum; Therefore the secondary user activity should be transparent to primary users. In the second proposed approach, all the users are treated equally and should limit their interference to their neighbors.

From the technological point of view there are still many challenges that need to be addressed until any one of these two approaches become applicable \cite{xG}, \cite{Implementationissues} and \cite{fundamentallimits}. In particular, both approaches require an acute interference management in order for the users to be able to coexist peacefully along side each other in a shared medium. In order to limit the interference to the primary users two main approaches has been suggested.

\begin{figure}
\centering
\includegraphics [width=5in]{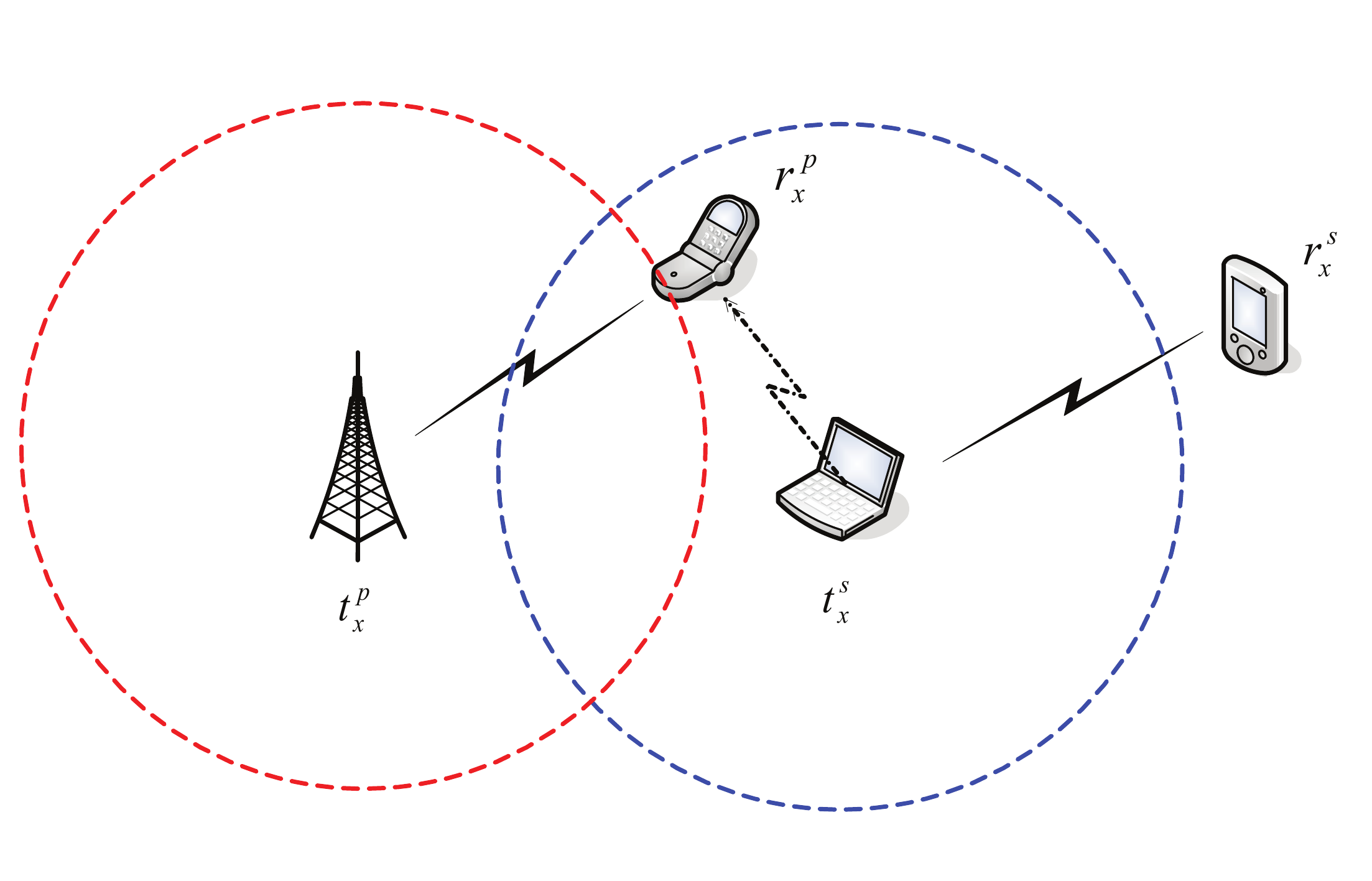}
\caption{Spectrum Sharing}
\label{fig:SpectrumSharing}
\end{figure}

\begin{figure}
\centering
\includegraphics [width=5in]{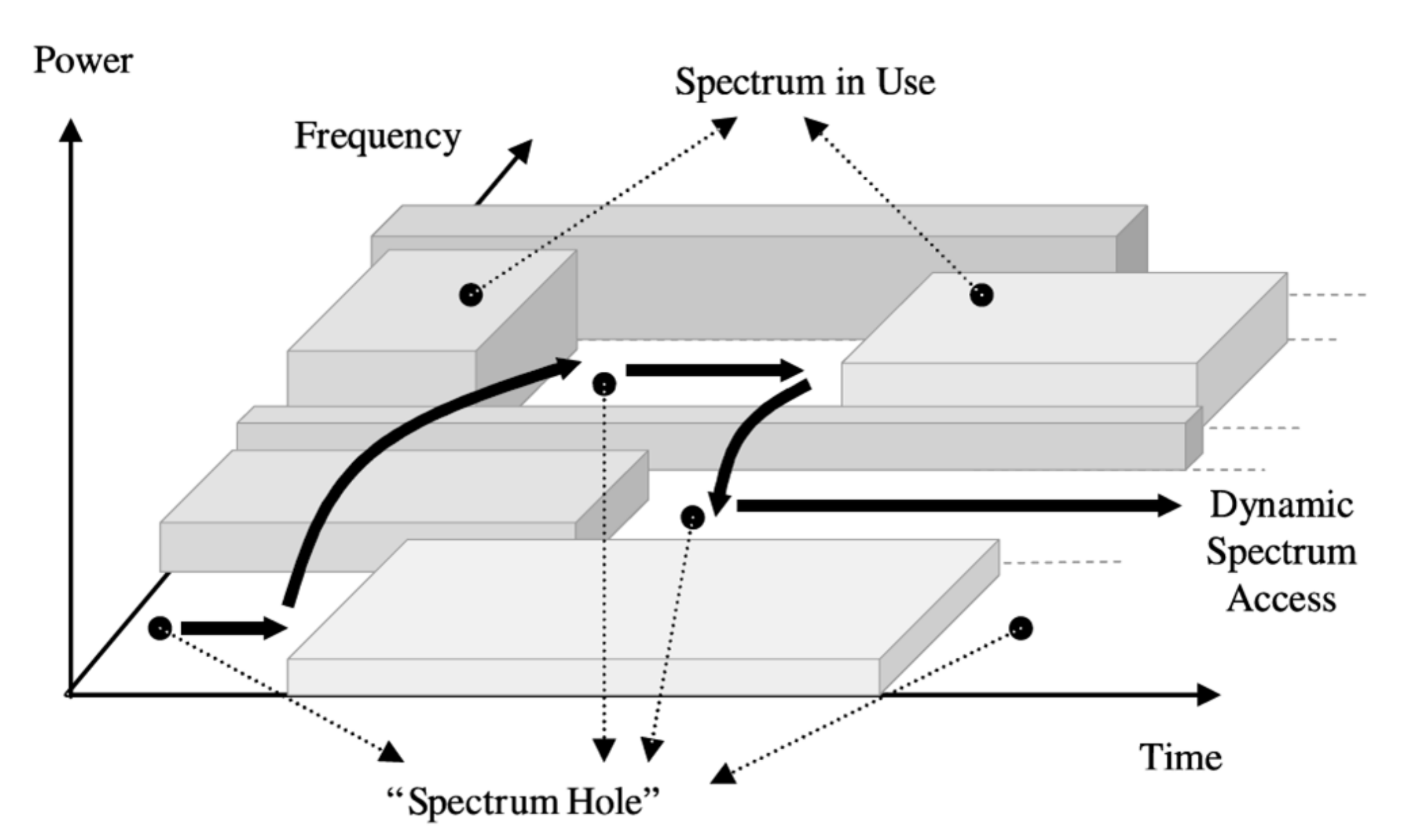}
\caption{Opportunistic Spectrum Access}
\label{fig:OpportunisticSpectrumAccess}
\end{figure}

The first one, {\it spectrum sharing}, is based on controlling the interference temperature at the primary users and makes use of ultra-wideband signaling. For example, secondary users could spread their power over a vast bandwidth to minimize the interference they cause to the primary users \cite{SpectrumSharing}; Essentially in this method, the secondary users  transmit their packages while the channels are occupied by the primary users but they schedule their transmissions such that the perceived interference at each primary receiver does not exceed the tolerable threshold, as shown in Fig. \ref{fig:SpectrumSharing}. The second strategy is called {\it opportunistic/dynamic spectrum access}, in which secondary users only make use of locally or temporally unused channels to transmit their data, as shown in Fig. \ref{fig:OpportunisticSpectrumAccess}. 
 Primary signal detection is of fundamental importance to this strategy; The performance of the sensing scheme and the detector characteristics highly affect the system performance. 
  In Chapters \ref{ThroughputAnalysisBasedonALOHAProtocol} and \ref{ThroughputAnalysisBasedonCSMA/CAProtocol} these effects will be studied in details. Since in this strategy secondary users presumably use the vacant channels they can transmit in higher power or bit rates compared with the first strategy; But they could only use the channel over a fraction of time or frequency.

The triumph of one strategy over the other is hard to justify due to many involved open problems. In this thesis we take the second approach (opportunistic spectrum accessing) and study the performance of the resulting cognitive radio system with randomized medium access protocols.




\section{Cognitive Radios}

The term {\it Cognitive radio} was first coined by Joseph Mitola III as a radio that is sufficiently intelligent about the radio resources and can identify the user communication needs in order to provide wireless services most appropriate to user needs \cite{MitolaI} \cite{MitolaII}. Mitola’s CR-1 cognitive radio prototype modeled a cognition cycle at the application layer. His research pointed to the potential use of cognitive radio technology to enable spectrum rental applications and create secondary wireless access markets \cite{Biologically}. A more common definition that restricts the radio's cognition to more practical sensory inputs is the FCC definition of cognitive radios as ``a radio that can change its transmitter parameters based on interaction with the environment in which it operates'' \cite{FCC 03-322}.

The main idea of cognition for a radios is to periodically search the spectrum for available opportunities (idle frequency bands), dynamically adopts the proper transmission policy (power, modulation scheme, \ldots) in order to avoid interference to other users. When we have multiple cognitive radios working together, we have a cognitive radio network.

\section{Motivation}

As mentioned in the previous section, the main  idea of cognitive networks is to allow unlicensed users (secondary users) to access the idle portion of the frequency spectrum without interfering (or, rather, doing so in a controlled manner) with the licensed users (primary users). Though the basic idea seems simple, this technology still faces many challenges (technological and regulatory). For instance, in order to explore opportunities and also limit the interference to the primary users, cognitive radios need to periodically scan the spectrum. One of the key challenges in this regard is the design of wide-band detectors \cite{Implementationissues}.

Considering present difficulties in processing multi-gigahertz bandwidth signals and reliable detection of primary users presence, the protocol introduced in this work is based on practical technology limits and assumes that each secondary user is capable of sensing and accessing only a limited number of frequency bands at a time. Hence, secondary users can only obtain partial information about the channel state which together with inherent hierarchy in accessing the channels, imposes substantial complications in identifying transmission opportunities and differentiates the cognitive medium access control (MAC) layer from the MAC layer in conventional radios.

One of the key issues in designing the cognitive MAC is how to decide which channel(s) to sense and how to share the idle channels among secondary users. A considerable amount of literature exists on cognitive MAC protocols \cite{POMDP} - \cite{Cross-LayerBasedOpportunisticMACProtocols}. In \cite{POMDP} and \cite{JointDesign}, an optimal strategy for dynamic spectrum access is developed by integrating the design of spectrum sensors at the physical layer with that of spectrum sensing and access policies at the MAC layer, based on a partially observable Markov decision processes (POMDPs) framework. Although this protocol can well utilize the residual spectral opportunities for a single secondary user, its implementation and analytical study of the network level performance (achievable secondary network throughput) is complicated. In \cite{abandwidthsharingapproach} the authors present an ad-hoc secondary medium access control (AS-MAC) protocol, which utilizes the resources available on the downlink time/frequency domain, for Mobile Communications (GSM) cellular network. Through simulation, it is shown that they are able to efficiently utilize the available resources, with a utilization factor from $75$ to $132$ percent, due to spatial reuse in the latter case. The authors of \cite{DynamicopenspectrumsharingMACprotocol} propose a dynamic open spectrum sharing (DOSS) MAC protocol, which is somewhat impractical as it requires three separate sets of transceivers to operate on the control channel, data channel, and busy-tone channel, respectively. In \cite{Cross-LayerBasedOpportunisticMACProtocols} the authors propose certain schemes, which integrate the spectrum sensing policy at the physical (PHY) layer with packet scheduling at the MAC layer. They analyze the throughput and the delay-QoS performances of the proposed schemes for the saturation network and the non-saturation network cases under random and negotiation-based channel sensing policies. However, their sensing strategy suffers from the fact that secondary users need to cooperate to sense the licensed channels through a pre-set control channel. Moreover, the secondary users pre-select the channels to access in the previous time slot, which faces the problem that there is no guarantee the channel occupancy states remain the same during two consecutive time slots. In addition, all the aforementioned work, except the last one, do not address the performance of cognitive MAC protocols at the network level and the consequences that each entity decision has on the performance of the others.

Furthermore, due to deployment difficulties, the cognitive schemes that make use of a central authority to coordinate the action of secondary users are less appealing. Therefore, recently, distributed techniques for dynamic spectrum allocation, where no central spectrum authority is required, are being widely studied. Though they are less efficient, decentralized approaches require much less cooperation. Some of the decentralized protocols require control channels as a common medium among locally adjacent users to negotiate the communication parameters in order to make the best use of the available opportunities \cite{DynamicopenspectrumsharingMACprotocol} \cite{Distributedcoordination}. These protocols face serious challenges: How to set up the control channel? What if the control channel is corrupted by interference? How to negotiate a transition to a new frequency?

Due to aforestated difficulties and the desired autonomous nature of secondary users, in this work we focus on randomized cognitive MAC protocols, which need no (or minimal) information to be exchanged among secondary users in order for them to take an appropriate actions. This fact rids us from the need of control channels and the complications that they incur.

\section{Outline}

In this thesis we propose and study cognitive MAC protocols for wireless ad-hoc networks, which integrate a randomized sensing scheme to decide on which channel to scan and Slotted ALOHA and CSMA/CA protocols to avoid collision among secondary users. Secondary users do not need any centralized controller or common control channel to coordinate their actions. Primary users are distributed according to a spatial Poisson process in the network cell and utilize their allocated frequency bands according to know statistics. Secondary users are allowed to opportunistically access the spectrum as long as they guarantee a certain probability of collision to the primary users. In order to meet the collision constraint, secondary users scan the frequency band(s) they intend to exploit for primary activity, and access it provided that it is idle.

In Chapter \ref{ThroughputAnalysisBasedonALOHAProtocol} we study the performance of the ALOHA-based cognitive MAC protocol. We first introduce the system model under study and analyze the throughput performance of the scheme with idealized system parameters. We then derive the optimal number of secondary users that can coexist in the network under the Slotted ALOHA MAC protocol for a special case. Second, under general system parameters we investigate the trade-off between the probability of opportunity-detection and the probability of collision. Utilizing the optimal detection radius we determine the total primary and secondary throughput. Furthermore, we establish a conjecture on the optimal number of secondary users and prove it for a special case.

In Chapter \ref{ThroughputAnalysisBasedonCSMA/CAProtocol} we consider the CSMA/CA-based cognitive MAC protocol. We first presume that  secondary user are only capable of sensing and accessing one channel at a time and compare the performance of the corresponding optimal scheme with the heuristic scheme introduced in Section \ref{ThroughputAnalysisBasedonALOHAProtocol}. Next, we extend the secondary users sensing and access capability to multiple channels at a time and study how much this extra feature can benefit the system performance. Finally we conclude the analysis by considering the effect of detection errors on the performance of the optimal cognitive MAC protocol. Chapter \ref{conclusion} concludes this work.
\chapter{Randomized Sensing Scheme Based on Slotted ALOHA Protocol}
\label{ThroughputAnalysisBasedonALOHAProtocol}

In this Chapter we analyze the performance of a randomized cognitive MAC protocol for wireless ad-hoc networks in which secondary users can only sense and access \emph{one} channel at a time. We assume secondary users employ Slotted ALOHA protocol to limit the interference among each other in the shared medium. According to this protocol whenever a user has a packet to transmit, it does so with probability $q$.

First, we analyze the performance of the system under asymptotic values for some network parameters. This corresponds to the analysis of the system in the Data Link Layer. Essentially we assume \emph{perfect} performance at the physical layer, i.e., no collision occurs at the physical layer. Next, we extend the analysis to lower layers and investigate the issues that arise from the distributed nature of the network. Specifically, we study the trade-off between the opportunity utilization and the probability of collision. Then, utilizing the optimal detection radius (which satisfies the collision constraint set by the primary network and achieves the maximum throughput for the secondary network), we determine the total primary and secondary throughput.

\section{The System Model}
\label{systemmodel}

\begin{figure}
\centering
\includegraphics [width=4.5 in]{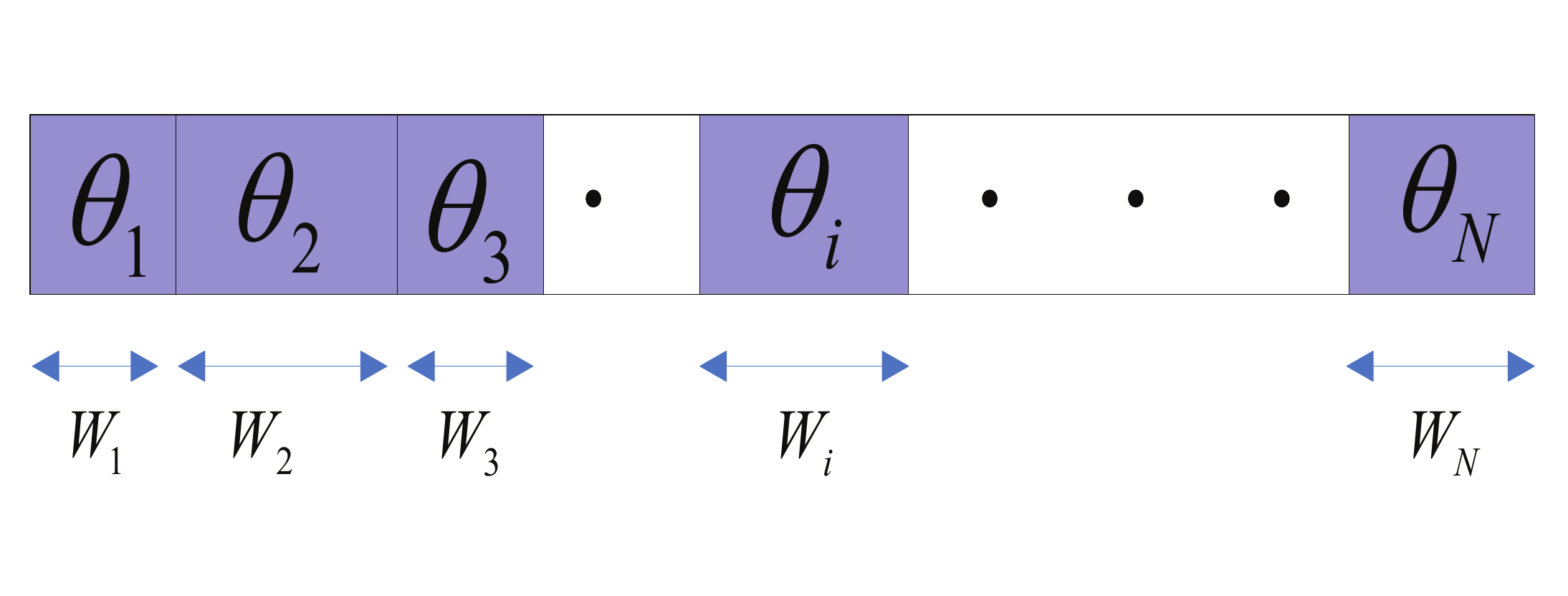} 
\caption{Primary network channel utilization statistics.}
\label{fig:Nchannels}
\end{figure}

Consider a frequency spectrum consisting of $N$ orthogonal channels with bandwidths $W_j$, $j \in \mathcal{N}$. Set $\mathcal{N} = \{1,\ldots,N\}$ represents the set of channel indices. The total bandwidth available to primary users is $\mathcal{W} = \sum_{j\in \mathcal{N}} W_j$. Each channel is licensed to a time-slotted primary network. The spectrum occupancy statistics of channels are independent of each other and follow a Bernoulli distribution with probability $\theta_j$ of being active for each channel $j \in \mathcal{N}$, Fig \ref{fig:Nchannels}.

Assuming a fixed Physical Layer scheme in all channels, the rate at which users communicate is proportional to their available bandwidth, i.e.

\begin{equation}
\label{proportionality}
C_j = a \cdot W_j
\end{equation}
Where  the proportionality constant $a$ depends on the PHY signaling scheme. Therefore, from the perspective of the data-link layer, the expected secondary network throughput equals

\begin{equation}
\label{primarythroughput}
C_p = \sum_{j=1}^N \theta_j\cdot C_j,
\end{equation}
whereas the total supported throughput in the network is potentially as in \ref{achievablerate}.

\begin{equation}
\label{achievablerate}
C_{max} = \sum_{j=1}^{N} C_j = a \cdot \mathcal{W}
\end{equation}

\begin{figure}
\centering
\includegraphics [width=5 in]{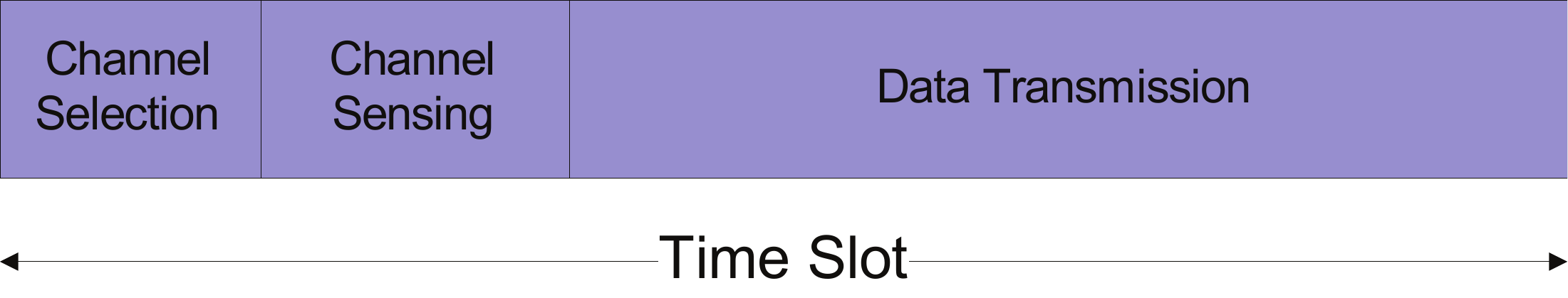}
\caption{The time-slot structure.}
\label{fig:timeslot}
\end{figure}

In order to utilize the spectrum more efficiently, the frequency channels are also made available to an unlicensed (secondary) network comprised of $M$ secondary users who seek opportunities to access the vacant frequency bands. The spectrum occupancy statistics of primary channels are assumed to be available to the secondary network. 
The secondary network is time-slotted and synchronized with primary network clock \footnote{We can assume both primary and secondary networks use the same clock source.}. We further assume that the secondary users operate under heavy traffic model, i.e. they always have packet in their queue to transmit. At the beginning of a time slot, each secondary user decides on which channel to sense according to a sensing scheme described later. Upon the perception of an opportunity \footnote{\emph{Opportunity} will be defined in following sections.} in that specific channel, the secondary user transmits a packet with probability $q$, in order to reduce the interference among secondary users (Slotted-ALOHA protocol \footnote{Alternatively, we can assume that data traffic arrives at each secondary transmitter in an i.i.d fashion with probability $q$ at the beginning of each time slot and gets transmitted upon reception.} ). However, if the channel is detected busy, the secondary user postpones the transmission of the data to the next time slot. The basic time slot structure is illustrated in Figure \ref{fig:timeslot}.

One of the challenges in establishing connection between a secondary transmitter and a receiver is synchronization \cite{Spectrumpooling}, \cite{Synchronizationalgorithms}. Since synchronization is out the scope of this work, for the sake of simplicity, we assume that the secondary transmitters and the receivers are perfectly synchronized, i.e. the secondary receiver knows in which channel the corresponding secondary transmitter is attempting to establish a connection at any time.

\begin{figure}
\centering
\includegraphics [width=6in]{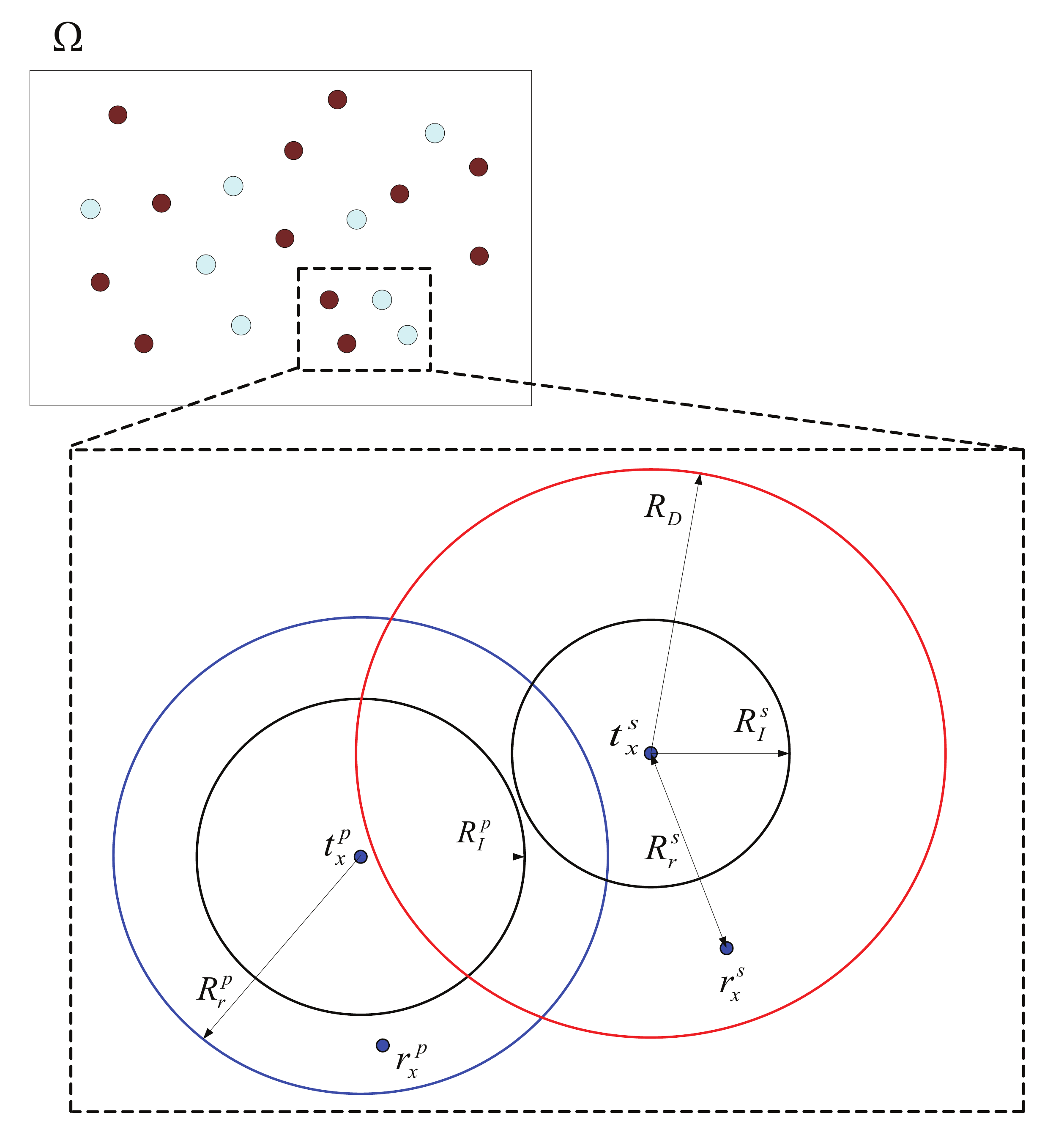}
\caption{Notations}
\label{fig:Notations}
\end{figure}

Primary users are distributed in the network cell, $\Omega$, according to two dimensional Poisson point process with density $\lambda$. Primary transmitters randomly pick a destination among the primary users located within their transmission range of radius $R_r^p$. 
To simplify the equations we assume that the secondary receivers are located at a fixed distance, $R_r^s$, away from the corresponding secondary transmitter. Secondary users are equipped with {\it perfect detectors} that can detect the primary signals without error within radius $R_D$. Moreover, secondary users are only cable of sensing and accessing a single channel at a time \footnote{We assumed a isotropic (spherical) model for signal propagation.}. In Section \ref{ThroughputAnalysisBasedonALOHAProtocol}-\ref{cellbasednetwork} we will show that even perfect detection does not guarantee collision free coexistence between primary and secondary users. Active receivers perceive interference from (non-designated) primary and/or secondary transmitters which are within $R_I^p$ and $R_I^s$ radii about them respectively. In Fig. \ref{fig:Notations} we illustrate the definition of the network parameters graphically and in Table \ref{tab:SystemParameters} summarize the system parameters.

\begin{table}
\caption{System Parameters}
\centering
\begin{tabular}{|c|c|}
\hline
Parameter & Definition\\
\hline
$\Omega$ & Network cell \\
$N$ & Number of primary frequency bands \\
$M$ & Number of secondary users \\
$(.)^p$ & Parameter associated with primary users\\
$(.)^s$ & Parameter associated with secondary users\\
$R_I$ & Interference range\\
$R_r$ & Transmission range\\
$R_D$ & Perfect detection range\\
$\lambda$ & Primary users density \\
$\theta_j$ & Probability of channel $j$ being busy\\
$q$ & Probability of packet transmission for secondary users\\
$W_j$ & Channel $j$th bandwidth\\
$C_j$ & Capacity of $j$th channel \\
$C_{max}$ & Maximum achievable rate in the whole network \\
$\rho$ & Primary users' utilization efficiency \\
$\mathbb{C}_j$ & Normalized data rate at channel $j$\\
\hline
\end{tabular}
\label{tab:SystemParameters}
\end{table}

Now we define what we mean by the sensing scheme in the most general case. Let $\mathcal{G} := \{G_1,\cdots,G_\kappa\}$ be the set of all sensing actions available to secondary users and $\kappa = \left|{\mathcal{G}}\right|$ be the cardinality of the action set. Assuming secondary users are capable of sensing $S$ channels at a time ($1 \leq S \leq N$), $G_i$ corresponds to a specific subset of $\mathcal{N}$ of the size $S$, i.e. for $i \in \{1, \cdots, \kappa\}$, $G_i \subseteq \mathcal{N}$, $\left|G_i\right| = S$. For $S = 1$, $G_i$ simply represents the channel $i$. Based on the specifications of the detector the value of $\kappa$ can vary from $N$ to ${N\choose S}$. We define the secondary users randomized \emph{sensing scheme} as the probability measure $(\mathcal{G},\mathbb{R},P)$, i.e.

\begin{eqnarray}
\label{defsensscheme}
\begin{split}
P_i &= P(G_i) \qquad i \in \{1, \cdots, \kappa\} \\
P_{i} &\geq 0 \qquad\qquad i \in \{1, \cdots, \kappa\} \\
\sum_{i = 1}^{\kappa} &P_{i} = 1
\end{split}
\end{eqnarray}
where $P_i$ denotes the probability that a secondary user chooses the group/channel $i$ to sense. In the next section we determine the throughput performance of the randomized sensing schemes in the data-link layer.

\section{Asymptotic Scenario}
\label{idealscenario}

In this section we consider the case in which $R_I^p$, $R_I^s$ and $R_D$ tend to infinity and determine the data-link layer performance of the randomized sensing scheme based on the ALOHA protocol. The immediate consequence of $R_I$ going to infinity is that any attempt by a particular transmitter to communicate in an already occupied channel (it could be occupied by either a primary or a secondary user pair) results in collision or packet loss for the active receiver. In other words, in this scenario, one and only one primary user will access each frequency band at a time. Furthermore, assuming $R_D$ going to infinity leads to the accurate detection of spectrum opportunities for secondary users, i.e. each secondary user can tell if there is any active primary user in a channel with absolutely no error.

As mentioned earlier, the performance metric of interest to us is the total secondary network throughput, i.e., the amount of data that is {\it successfully} transmitted per unit time, given that the interference caused to the primary users does not exceed an acceptable level.

The infinite detection radius for secondary users guarantees collision free communication for the primary network. Therefore, the expected total throughput of the primary network remains unchanged in the presence of secondary users as in \eqref{primarythroughput}. In the following section we derive the throughput of the secondary network assuming an arbitrary sensing scheme and present the numerical results based on a heuristic sensing scheme.

\subsection{Secondary Network throughput}

Assume there are M secondary user pairs competing for residual spectrum resources in the network. At the beginning of each time slot, each secondary user decides on which channel to sense according to the sensing scheme defined in \eqref{defsensscheme} for $S = 1$ and transmit its data with probability $q$. First, we derive the throughput of a single secondary user, for $N > 1$ and $q <1$, provided that there are exactly $k$ secondary users active in the network. Let $\overline{\theta_j}=1-\theta_j$ denote the probability that channel $j$ is idle, which is also a measure of the availability of channel $j$ to secondary users.

\begin{equation}
\label{singleSU}
C_{\textrm{single SU} \mid k} = \sum_{j=1}^{N} \overline{\theta_j}.C_j P_j(1-P_j)^{k-1}
\end{equation}
Thus the expected total throughput of the secondary network is given by

\begin{align}
C_s &= \sum_{i=1}^M {M\choose i} q^i(1-q)^{M-i}iC_{\textrm{single SU}\mid i} \nonumber \\
    &= q \sum_{j=1}^{N} \overline{\theta_j}.C_j P_j \sum_{i=1}^N i{M\choose i} \bigl(q(1-P_j)\bigr)^{i-1}\bigl(1-q\bigr)^{M-i} \nonumber \\
    &= qM \sum_{j=1}^{N} \overline{\theta_j}.C_j P_j (1-P_j)(1-qP_j)^{M-1} \label{secondarythroughput}
\end{align}

We observe that for the special cases of $N = 1$ and/or $q = 1$ Equation \eqref{secondarythroughput} does not hold. For $N = 1$, we get $P_1 = 1$. Therefore, by Equation \eqref{singleSU}, the throughput of a single secondary user given that $k$ secondary users are transmitting would be:

\begin{equation}
\label{singleSUcase1}
\begin{split}
C_{\textrm{single SU}\mid k} &= \sum_{j=1}^{N} P_j(1-P_j)^{k-1} \overline{\theta_j}.C_j \\
                &= P_1(1-P_1)^{k-1} \overline{\theta_1}.C_1
\end{split}
\end{equation}
The only case in which $C_{single SU\mid k}$ admits a non-zero value is $k = 1$, which results in $C_{\textrm{single SU}\mid k} = \sum_{j=1}^N \overline{\theta_j}C_j$, i.e. only one secondary user is active in the network. Following the same approach as in Equation \eqref{secondarythroughput} we get the total secondary throughput to be $M (\sum_{j=1}^N \overline{\theta_j}\cdot C_j) q(1-q)^{M-1} $. Similarly, for the case of $q = 1$, we have that \emph{all} the secondary users will transmit their packets in every time slot. Therefore, each secondary user achieves a non-zero throughput if it is the only user to choose a specific channel to sense. This amounts to a total secondary throughput of $M \sum_{j=1}^N \overline{\theta_j}\cdot C_j P_j(1-P_j)^{M-1}$. In the next section we introduce a heuristic sensing scheme and study its performance. Note that equation \eqref{secondarythroughput} is not concave (or quasi-concave) in $P$, hence, we can not obtain a globally optimal sensing scheme.

\subsection{A Suboptimal Sensing Scheme}
\label{suboptimalsensingscheme}

Since secondary users can only sense one channel at a time, they have partial knowledge of the channel state. According to Section \ref{ThroughputAnalysisBasedonALOHAProtocol}-\ref{systemmodel}, we define the {\it sensing scheme} as a probability measure $P$ on set $\mathcal{G}$, i.e. $(\mathcal{G},\mathbb{R},P)$. Since in this section we assume that secondary users are only capable of sensing a single channel each time, the set $\mathcal{G}$ is simply the set of all primary channel. Each secondary user selects a channel to sense according to probability measure $P$, i.e., the probability that a secondary user selects the $j^{th}$ channel to sense is equal to $P_j = P(G_j)$. By exploiting such random {\it sensing scheme}, unlike deterministic policies, we give the secondary users a degree of freedom in seeking spectrum opportunities and therefore enable the secondary network to achieve a higher expected throughput.

In this section we introduce a simple and intuitive suboptimal sensing scheme. We introduce the channel probability assignment as
\begin{subequations}
\label{searchprobability}
\begin{align}
P_j :&= \frac{\overline{\theta_j}\cdot C_j}{\sum_{i=1}^N \overline{\theta_i}\cdot C_i} \\
    &= \frac{1-\theta_j}{1-\rho} \cdot \mathbb{C}_j, \qquad i \in \mathcal{N}
\end{align}
\end{subequations}
Where $\rho := C_p/C_{max}$ represents the primary users {\it utilization efficiency} and $\mathbb{C}_j := C_j/C_{max}$ is the normalized data rate at channel $j$.

By this probability association, regarding the channel selection, each secondary user gives higher weight to the channels with higher expected throughput, which means each secondary user also works in favor of his own throughput ({\it win-win strategy}). In other words, it is more likely for a secondary user to choose a channel which has higher data rate and is less utilized with respect to primary network's utilization efficiency factor $\rho$. Furthermore, secondary users can assign the channel sensing probabilities arbitrarily and without any knowledge about other secondary users which serves the distributed nature of secondary network better. Also it is clear that secondary users have no incentive to deviate from this scheme in order to gain possible throughput boost. This assignment (or strategy) is also \emph{Pareto efficient}, i.e. there exist no other strategies that improves a particular users throughput and does not leave the others worse off. Therefore, throughout this chapter we assign the channel sensing probabilities as \eqref{searchprobability}. Also it can be easily proven that this probability assignment is the Nash equilibrium when number of secondary users tend to infinity \cite{CMACEEC}.

\begin{figure}
\centering
\includegraphics[width=5.5in]{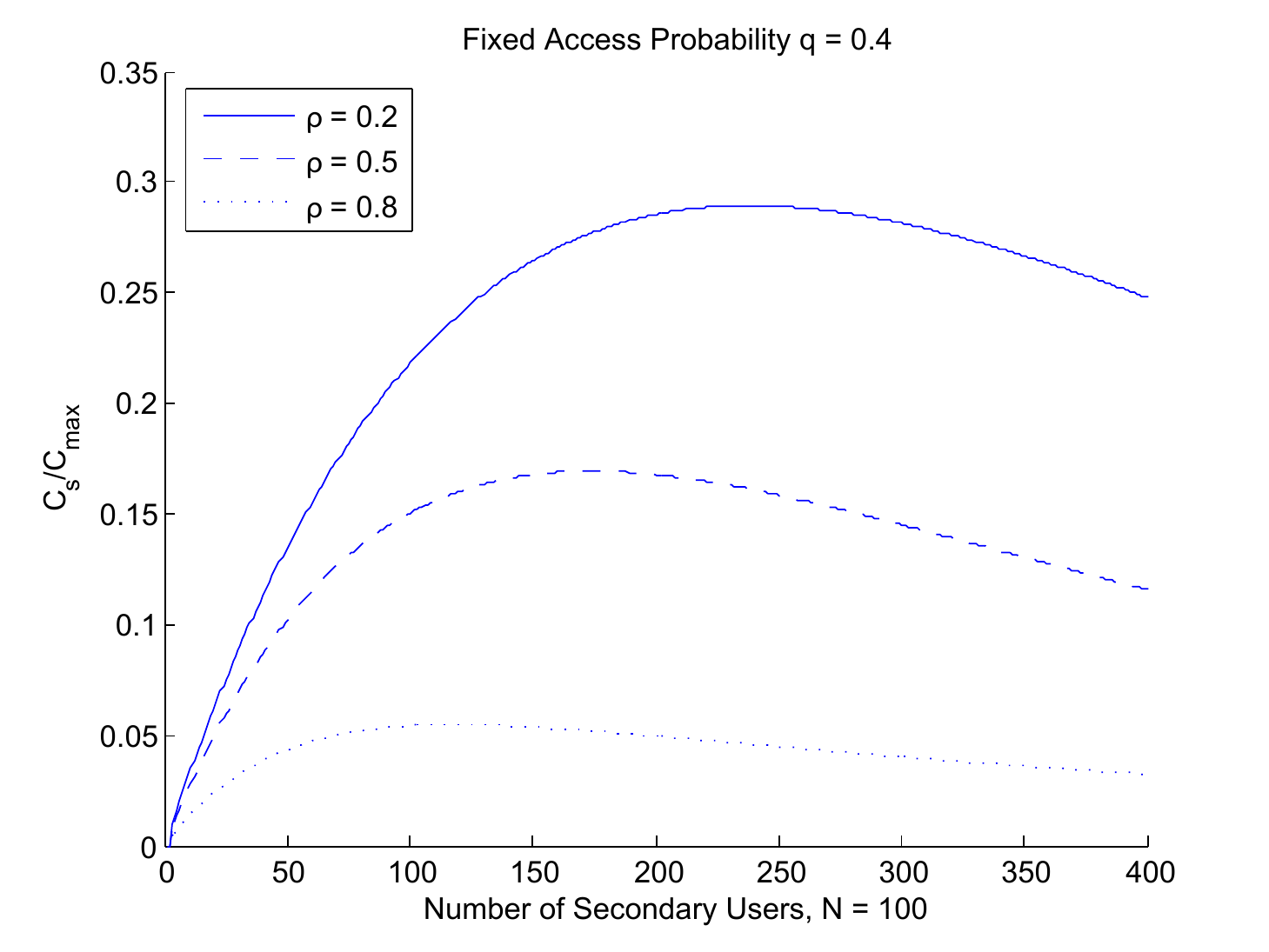}
\caption{Normalized secondary network throughput variation with respect to the number of secondary users for $q=0.4$, $N=100$ and three channel utilization statistics; $\rho = 0.2$ (Rarely Utilized), $\rho = 0.5$ (Fairly Utilized), $\rho = 0.8$ (Highly Utilized).}
\label{fig:Moptimization}
\end{figure}

By substituting \eqref{searchprobability} into \eqref{secondarythroughput}, we derive the secondary network throughput as:

\begin{equation}
\label{secondarythroughput2}
C_s = qMC_t \sum_{j=1}^N P_j^2(1-P_j)(1-qP_j)^{M-1}
\end{equation}
where $C_t = \sum_{j=1}^N \overline{\theta_j}\cdot C_j$ is the maximum achievable rate for secondary network. Fig. \ref{fig:Moptimization} depicts the variation of the normalized secondary network throughput with respect to the number of secondary users for fixed $q = 0.4$. As shown in Fig. \ref{fig:Moptimization} as excessive number of secondary users try to access the spectrum the system performance deteriorates and total throughput declines.

In order to gain an insight on the relation between optimal number of secondary users and the system parameters, lets assume $\theta_j = \theta$ and $C_j = C$ for all $j \in \mathcal{N}$. By doing so, we get $P_j = 1/N$ for all $j \in \mathcal{N}$ and $C_s = qMC_t \frac{1}{N}(1-\frac{1}{N})(1-\frac{q}{N})^{M-1}$. It is easy to prove that $C_s$ is quasi-concave over $M$ and $M^* = -1/\ln(1-q/N)$ which for large enough $N$ can be approximated as $M^* \approx N/q$. This suggests that the optimal number of secondary users which can co-exist in the secondary network depends on the number of the channels (or the system \emph{spectrum diversity}) and how remissive secondary users are, i.e. $1/q$.

\section{General Scenario: Cell Based Network}
\label{cellbasednetwork}

In this section we determine the system throughput, considering the spatial constellation of users in the network cell. Note that the $R_I < \infty$ and $R_D < \infty$ are the implicit assumptions in this section. $R_I < \infty$ corresponds to the inherent spatial diversity in wireless networks. And $R_D < \infty$ corresponds to the local and distributed knowledge of the secondary users regarding the legacy network.

As mentioned in Section \ref{ThroughputAnalysisBasedonALOHAProtocol}-\ref{systemmodel}, at the beginning of each time slot, a data communication by primary users is taken place in channel $j$ with probability $\theta_j$. Therefore according to coloring and displacement theorems for Poisson point processes, \cite{poissonprocesses} and \cite{mixedpoissonprocesses}, the primary transmitters and receivers active in channel $j$ both form a two dimensional Poisson point process with rate $\lambda \theta_j$. Note that due to limited interference radius for both primary and secondary transmitters, there can be more than one active primary or secondary user in network cell $\Omega$ in each spectrum channel. In addition, there are $M$ secondary users uniformly distributed in the network cell.

At the beginning of each time slot secondary users with data to transmit choose a channel to sense according to the randomized sensing scheme defined by \eqref{searchprobability}. The channel is perceived as an \emph{opportunity} if no primary receiver is detected within the interference radius of the secondary transmitter, $R_I^s$, with high enough probability. Although secondary users can detect perfectly within radius $R_D$, the detection of primary receiver is not error free. This error is caused by the distributed nature of the network. Therefore, by assuming perfect detection for secondary users we focus our study only on the issues that distributed nature of the ad-hoc networks imposes on the performance of the system. As it is shown in Section \ref{ThroughputAnalysisBasedonALOHAProtocol}-\ref{cellbasednetwork}-\ref{collisionprob} the probability of error in detecting the primary receivers depends on the detection radius. Therefore, the detection radius of the secondary users must be chosen wisely in order to enable them to communicate without violating the interference constraint imposed by the primary network.

A packet transmission is considered \emph{successful} if there is no primary or secondary transmitters present within radii $R_I^p$ and $R_I^s$ of the secondary receiver, respectively. Here we assume that no acknowledgment is required to complete a packet transmission, i.e. secondary transmitters do not retransmit the data which is lost in the channel (for instants, on-line video or audio streaming and gaming). In Section \ref{ThroughputAnalysisBasedonALOHAProtocol}-\ref{cellbasednetwork}-\ref{successfultrans} we derive the expression for the probability of successful transmission.

\begin{figure*}
\centering
\includegraphics[width=6in]{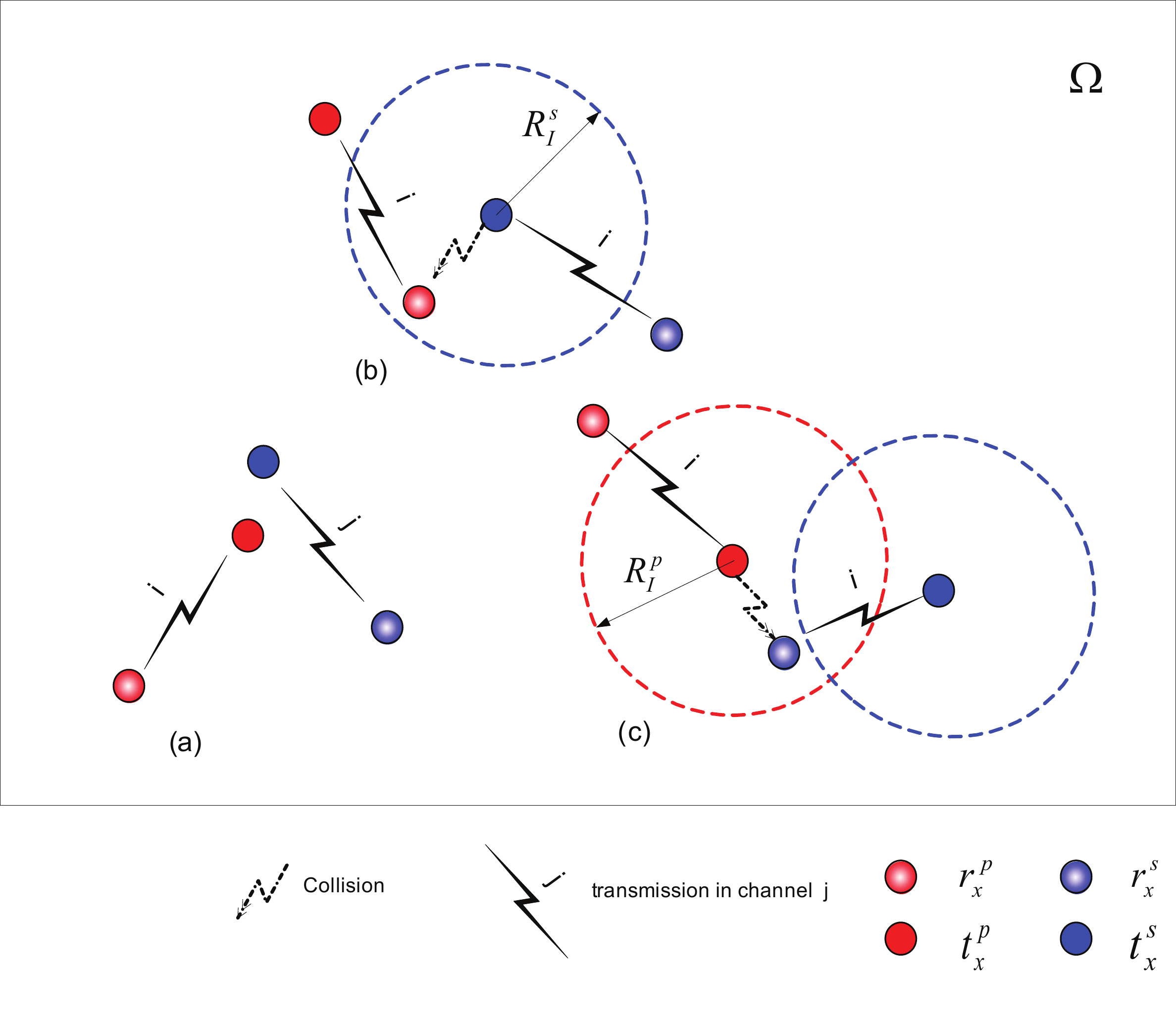}
\caption{An illustration of a successful and interfering transmissions (a) both primary and secondary users pair successfully communicate (b) secondary user pair successfully communicates but interferes with primary communication (c) primary user pair successfully communicates but interfere with secondary user's communication.}
\label{fig:illustration}
\end{figure*}

\begin{table}
\caption{Definition of the Successful transmission and the Spectrum Opportunity}
\label{tab:DefinitionoftheSuccessfultransmissionandtheSpectrumOpportunity}
\centering
\begin{tabular}{|c|c|}
\hline
Terminology & Definition\\
\hline
Successful transmission & No $t_x^p$ or $t_x^s$ present within $R_I^p$ and $R_I^s$\\
& range of an active receiver\\
Spectrum Opportunity & No primary receiver detected within interference \\
& radius of a SU with high probability\\
\hline
\end{tabular}
\end{table}

Fig. \ref{fig:illustration} illustrates three scenarios of successful and interfering data communication. In Fig. \ref{fig:illustration}(a), primary and secondary user pairs communicate in different spectrum channels and since we assumed spectrum channels are orthogonal to each other both pairs communicate successfully. However in Fig. \ref{fig:illustration}(b), the secondary transmitter miss detects the opportunity and interferes with the signal reception of the primary receiver and causes a packet loss. On the other hand, since there is no active primary or secondary transmitters in vicinity of the secondary receiver, the secondary receiver successfully receives its data. In Fig. \ref{fig:illustration}(c), primary user pair can communicate without interference. The secondary transmitter detects a spectrum opportunity and starts transmission, however, data communication of primary users interferes with the reception of the secondary receiver.

\subsection{Primary Network Throughput and Probability of Collision}
\label{collisionprob}

As mentioned in Section \ref{ThroughputAnalysisBasedonALOHAProtocol}-\ref{cellbasednetwork} in order to abide by the collision constraint of primary network, secondary transmitters must transmit if and only if there is no primary receiver active within radius $R_I^s$ of them. Since detecting primary receiver directly is impossible without receiver's cooperation, secondary transmitters need to base their decision on the existence of primary signal within their detection region, i.e. whether there is primary transmitter within radius $R_D$. In other words, in practice, a channel is perceived as an \emph{opportunity} if there is no primary transmitters detected within $R_D$ radius of a secondary transmitter. Nonetheless, there is a trade off between the probability of collision and the probability of overlooking a spectrum opportunity.

\begin{figure*}[!t]
\centering
\includegraphics[height=4 in]{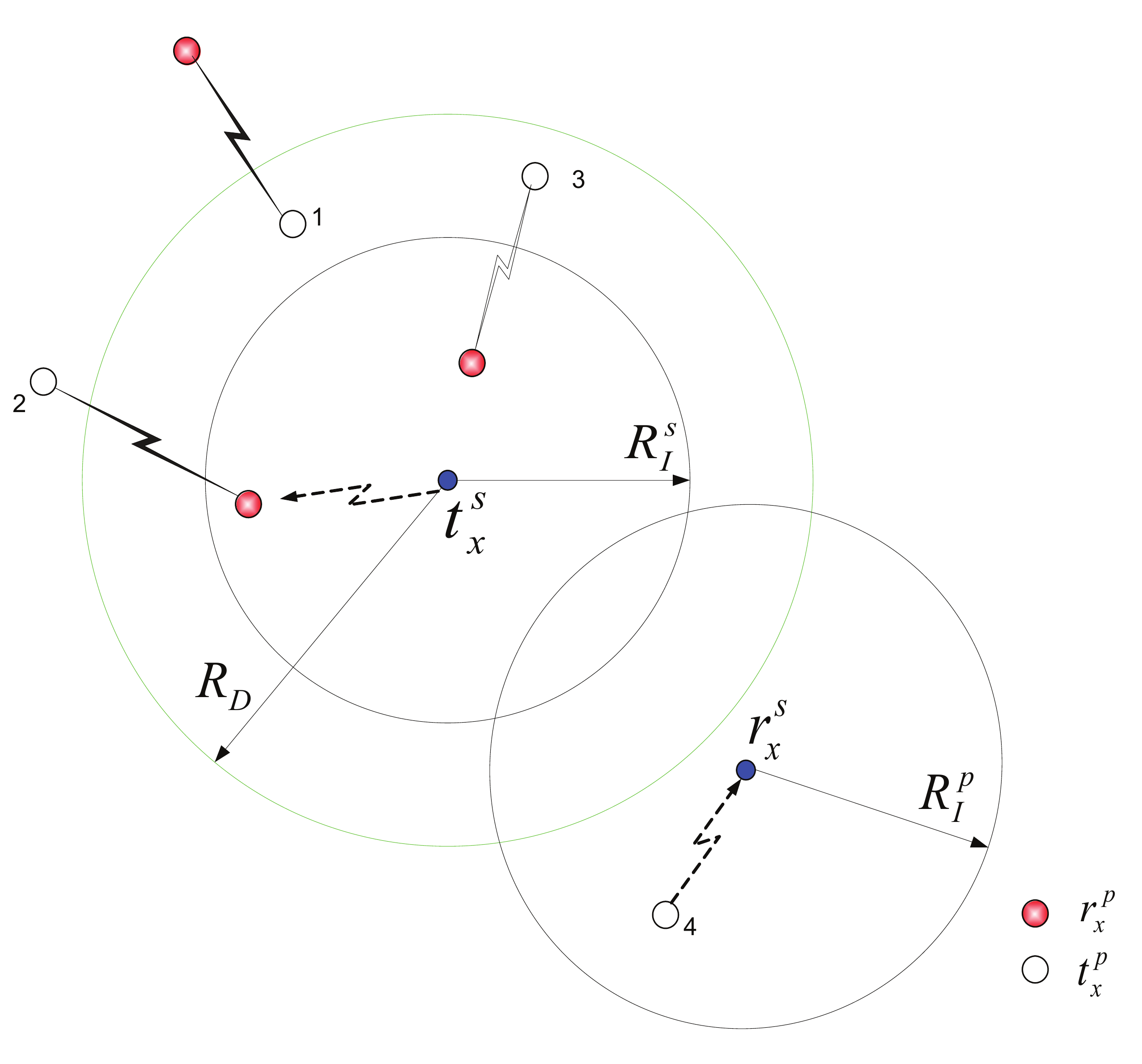}
\caption{An illustration of the opportunity detection by secondary users}
\label{fig:collisopm}
\end{figure*}

By choosing $R_D = R_I^s + R_r^p$ we can make the probability of collision \emph{equal to zero}  in expense of increasing the probability of overlooking a spectrum opportunity and degrading secondary network throughput. As shown in Fig. \ref{fig:collisopm} with the given detection radius $R_D$, the secondary transmitter can successfully avoid collision to the primary receiver by detecting Primary Transmitter $3$, but it can not detect Primary Receiver $2$ since it's corresponding transmitter falls outside of detection range which results in collision. However, by detecting Primary Transmitter $1$, the secondary transmitter refrains from transmission although the corresponding primary receiver falls outside the interference region of the secondary transmitter. Moreover, due to spatial distribution of primary and secondary users in the network this effect is aggregated by the number of active secondary user pairs in the same channel.

Therefore, secondary transmitters need to select $R_D$ large enough such that it guarantees the primary network collision constraint as well as refraining secondary network from overlooking too many opportunities.

\begin{table}
\caption{Events Definition}
\label{tab:EventsDefinition}
\centering
\begin{tabular}{|c|c|}
\hline
Event & Definition\\
\hline
$E_x(y,r)$ & Existence of at least a user $y$ within $r$ range of user $x$\\
$\mathfrak{C}_{R}(r,\varphi)$ & Circle centered at polar coordinates $[r,\varphi]$ with radius $R$\\
$\left|\mathcal{T}\right|$ & Lebesgue measure of set $\mathcal{T}$ \\
\hline
\end{tabular}
\end{table}

Let $E_x(y,r)$ be the event in which there exists at least one user of type $y$ within radius $r$ of user $x$. Also let $\mathfrak{C}_{R}(r,\varphi)$ denote a circle centered at polar coordinates $[r,\varphi]$ with radius $R$ and $\left|\mathcal{T}\right|$ denote the area of region $\mathcal{T}$. Based on the earlier discussion, \emph{collision} is the event in which there exists a primary receiver within the interference range of a secondary user while the secondary user can not detect a primary transmitter within its detection range. Consequently, we can compute the probability of collision in channel $j$ as follows:

\begin{equation}
\label{allcollisioninj}
\begin{split}
P_{c\mid j}& = \sum_{l=1}^M \sum_{k=1}^l \biggl[ \textrm{Pr}(\{ \text{l SU being active} \})  \\
& \cdot \textrm{Pr}(\{ \text{k out of l active SU choose channel j to transmit} \}) \\
& \cdot k \cdot \textrm{Pr}\bigl({\overline{E_{t_x^s}(t_x^p,R_D)}}\mid E_{t_x^s}(r_x^p,R_I^s)\bigr) \biggr] \\
& = \sum_{l=1}^M {M\choose l} q^l(1-q)^{M-l} \sum_{k=1}^l {l\choose k} P_j^k(1-P_j)^{l-k} k P_{cc\mid j} \\
& = P_{cc\mid j} \sum_{l=1}^M {M\choose l} q^l(1-q)^{M-l} l P_j
\end{split}
\end{equation}
which yields to:

\begin{equation}
P_{c\mid j} = MqP_jP_{cc\mid j}
\end{equation}
where $P_{cc\mid j}$ is the probability that the secondary transmitter wrongly perceives channel $j$ as an opportunity (i.e. there is no active primary transmitters in channel $j$ within $R_D$ radius of the secondary transmitter) while there is at least one active primary receiver within its $R_I^s$ radius. After some mathematical manipulation, we derive the $P_{cc\mid j}$ as \eqref{collisioninj_2}.

\begin{subequations}
\label{collisioninj}
\begin{gather}
P_{cc\mid j} = \textrm{Pr}\bigl({\overline{E_{t_x^s}(t_x^p,R_D)}}\mid E_{t_x^s}(r_x^p,R_I^s)\bigr) \\
         = \frac{\textrm{Pr}\bigl({\overline{E_{t_x^s}(t_x^p,R_D)}}\cap E_{t_x^s}(r_x^p,R_I^s)\bigr)}{E_{t_x^s}(r_x^p,R_I^s)}  \nonumber\\
         = \frac{\textrm{Pr}\bigl({\overline{E_{t_x^s}(t_x^p,R_D)}}\bigr)}{1-\textrm{Pr}\bigl({\overline{E_{t_x^s}(r_x^p,R_I^s)}}\bigr)}\biggl[1-\textrm{Pr}\bigl({\overline{E_{t_x^s}(r_x^p,R_I^s)}}\mid {\overline{E_{t_x^s}(t_x^p,R_D)}}\bigr)\biggr] \nonumber \\
         = \frac{\textrm{Pr}\bigl({\overline{E_{t_x^s}(t_x^p,R_D)}}\bigr)}{1-\textrm{Pr}\bigl({\overline{E_{t_x^s}(r_x^p,R_I^s)}}\bigr)} \biggl[1-\sum_{k=0}^\infty \biggl(\textrm{Pr}\bigl({\overline{E_{t_x^s}(r_x^p,R_I^s)}}\mid E_{t_x^s}^k(t_x^p,R_I^s+R_r^p),{\overline{E_{t_x^s}(t_x^p,R_D)}}\bigr) \cdot  \nonumber \\
          \textrm{Pr}\bigl(E_{t_x^s}^k(t_x^p,R_I^s+Rr)\mid {\overline{E_{t_x^s}(t_x^p,R_D)}}\bigr)\biggr)\biggr]  \label{collisioninj_1}\\
         = \frac{\exp(-\lambda \theta_j\pi R_D^2)}{1-\exp(-\lambda \theta_j\pi R_I^{s^2})}\biggl[1-\sum_{k=0}^\infty S^k \frac{(\lambda \theta_j\pi\tilde{S})^k}{k!}\exp(-\lambda \theta_j\pi\tilde{S})\biggr]  \nonumber \\
         = \frac{\exp(-\lambda \theta_j\pi R_D^2)}{1-\exp(-\lambda \theta_j\pi R_I^{s^2})} \left[1-\exp\bigl(-\lambda \theta_j\pi\tilde{S}(1-S)\bigr)\right] \label{collisioninj_2}
\end{gather}
\end{subequations}
where

\begin{subequations}
\label{Sandtilde(S)}
\begin{gather}
\tilde{S} = (R_I^s+R_r^p)^2-R_D^2 \\
S = \frac{2}{\tilde{S}} \int_{r=R_D}^{R_I^s+R_r^p} \frac{\left|\mathfrak{C}_{R_r^p}(r,0)-\mathfrak{C}_{R_I^s}(0,0)\right|}{\pi (R_r^p)^2}~rdr
\end{gather}
\end{subequations}
and $\left|\mathfrak{C}_{R_I^s}(0,0)-\mathfrak{C}_{R_r^p}(r,0)\right|$ is the area inside a circle with radius $R_I^s$ and outside another circle distance $r$ apart with radius $R_r^p$. In Equation \eqref{collisioninj_1} we make use of the fact that if there is an active primary receiver in $R_I^s$ radius of the secondary transmitter, its associated (primary) transmitter should be somewhere within $R_I^s+R_r^p$ radius of the secondary transmitter. Also note that each secondary transmitter causes collision to primary receivers independent of one another.

\begin{figure}
\centering
\includegraphics [width=5.5 in]{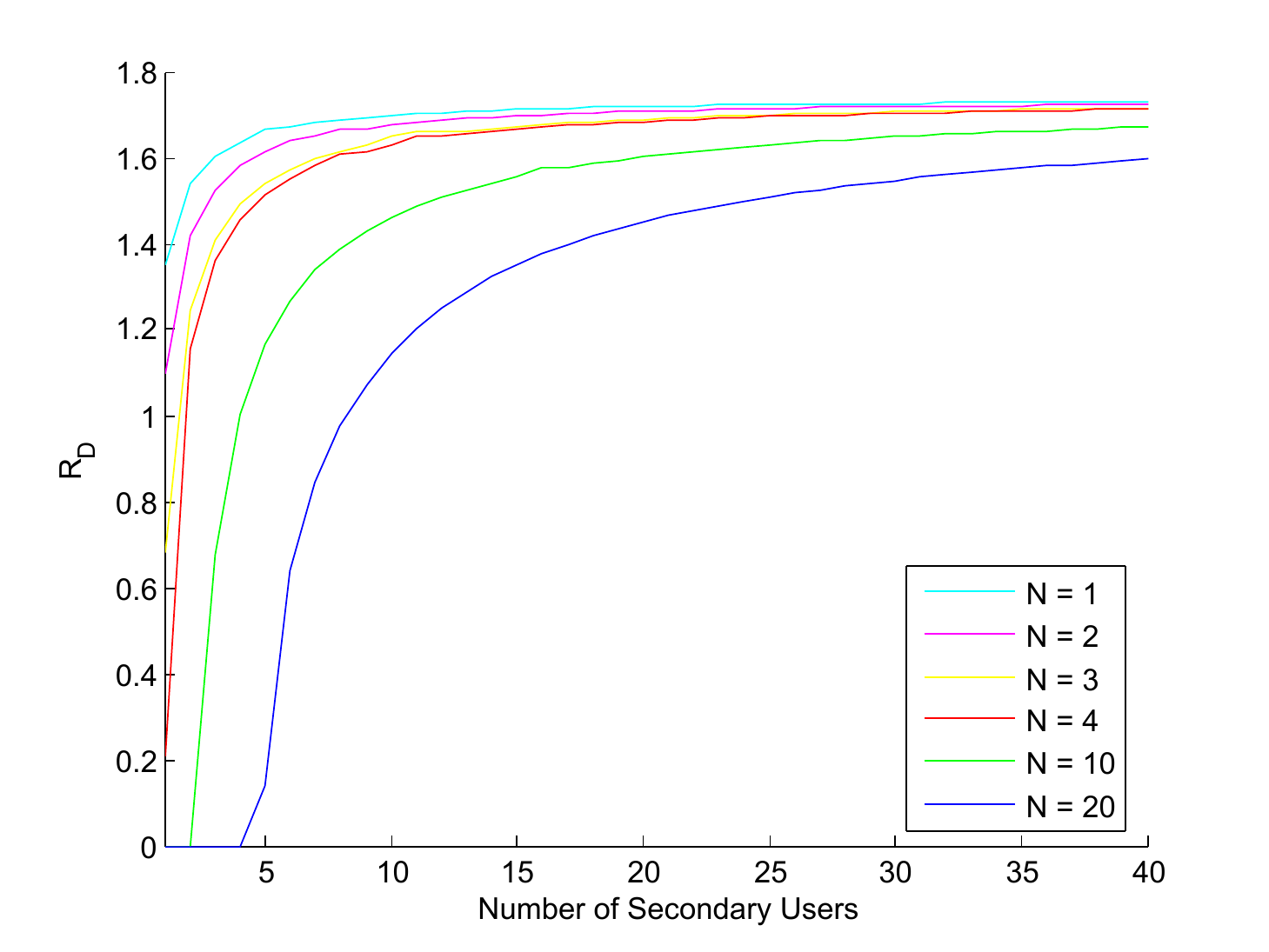}
\caption{Variation of the detection radius with respect to number of secondary users and number of primary frequency channel. The system parameters are set as: $R_I^{s,p}=R_r^{s,p}=1$, $\xi = 0.2$, $q = 0.3$,$\lambda = \frac{1}{1.5^2}$}
\label{fig:RdVSM}
\end{figure}

In order to meet the primary network's average collision constraint (i.e. probability of collision in all channel not exceeding $\xi$) we need to satisfy:

\begin{subequations}
\label{PUcollisionConstraint}
\begin{align}
\frac{1}{N} \sum_{j=1}^N P_{c\mid j} & \leq \xi \\
or ~  \sum_{j=1}^N P_jP_{cc\mid j} & \leq \frac{N}{Mq} \xi
\end{align}
\end{subequations}
Which yields to the selection of optimal detection radius. Fig. \ref{fig:RdVSM} demonstrate the variation of the optimal detection radius with respect to the number of secondary users and the number of primary frequency channel. As it can be concluded from the Equation \eqref{PUcollisionConstraint}, the bigger $N$ and/or smaller $M$ results in smaller $R_D$. We can think of the ratio $N/M$ as a measure of congestion in the network. A bigger $N/M$ represents a bigger scatter (or diversity) of the secondary users in the frequency channels which brings about less collision with the primary users. As an special case, for $N = 20$, $R_D$ equals zero for $M \leq 4$ because $N$ is so bigger compared to $M$ that allows the secondary users to be able to meet the collision constraint even without sensing the spectrum.



\subsection{Secondary Network Throughput and Probability of Successful Transmission}
\label{successfultrans}

As mentioned in Section \ref{cellbasednetwork}, a transmission is considered successful if there is no primary and secondary transmitter present in $R_I^p$ and $R_I^s$ radii of an active secondary receiver, respectively. Therefore, the probability of the successful transmission for a single secondary user pair in channel $j$ given there are $K$ active secondary transmitter present in that frequency band is:

\begin{equation}
\label{probofsuccessfultrans}
P_{s \mid (j,K)} = \textrm{Pr}\bigl({\overline{E_{t_x^s}(t_x^p,R_D)}}\cap{\overline{E_{r_x^s}(t_x^p,R_I^p)}}\cap{\overline{E_{r_x^s}(t_x^s,R_I^s)}} \mid K \bigr)
\end{equation}
In which the first term ${\overline{E_{t_x^s}(t_x^p,R_D)}}$ indicates the event that the secondary transmitter detects an opportunity and initiates the transmission. The last two terms ${\overline{E_{r_x^s}(t_x^p,R_I^p)}}$ and ${\overline{E_{r_x^s}(t_x^s,R_I^s)}}$ indicate a collision (due to primary transmitter and other secondary transmitters) free transmission.

We can expand Equation \eqref{probofsuccessfultrans} as,

\begin{align}
\label{appencalculationPsj}
P_{s\mid (j,K)} &= \textrm{Pr}\bigl({\overline{E_{t_x^s}(t_x^p,R_D)}}\cap{\overline{E_{r_x^s}(t_x^p,R_I^p)}} \mid K \bigr) -
        \textrm{Pr}\bigl({\overline{E_{t_x^s}(t_x^p,R_D)}}\cap{\overline{E_{r_x^s}(t_x^p,R_I^p)}}\cap E_{r_x^s}(t_x^s,R_I^s) \mid K \bigr) \nonumber \\
        &= \textrm{Pr}\bigl({\overline{E_{t_x^s}(t_x^p,R_D)}}\cap{\overline{E_{r_x^s}(t_x^p,R_I^p)}}\bigr) - P_{\mathbb{I}}^{(j,K)} \nonumber \\
        &= exp\biggl(-\lambda \theta_j \Bigl[\pi(R_D^2+R_I^{p^2})-\left|\mathfrak{C}_{R_D}(0,0)\cap\mathfrak{C}_{R_I^p}(R_r^s,0)\right|\Bigr]\biggr) - P_{\mathbb{I}}^{(j,K)}
\end{align}
where
\begin{align}
\label{PI}
P_{\mathbb{I}}^{(j,K)} &= \textrm{Pr}\biggl({\overline{E_{t_x^s}(t_x^p,R_D)}}\cap{\overline{E_{r_x^s}(t_x^p,R_I^p)}}\cap \bigl[\bigcup\limits_{m=1}^{K-1}E_{r_x^s}^m(t_x^s,R_I^s)\bigr]\biggr) \nonumber \\
               &= \sum_{m=1}^{K-1} \textrm{Pr}\bigl({\overline{E_{t_x^s}(t_x^p,R_D)}}\cap{\overline{E_{r_x^s}(t_x^p,R_I^p)}}\mid E_{r_x^s}^m(t_x^s,R_I^s)\bigr)\cdot\textrm{Pr}\bigl(E_{r_x^s}^m(t_x^s,R_I^s)\bigr)  \nonumber \\
               &= \sum_{m=1}^{K-1} P_{\mathbb{II}}^{(m,j)} \cdot {K-1\choose m}\gamma^m(1-\gamma)^{K-1-m}
\end{align}

$\gamma = \frac{\left|\mathfrak{C}_{R_I}(.)\right|}{\left|\Omega\right|}$ denotes the ratio between interference region area to the network cell area \footnote{Which is typically small.}. Also notice that $1/\gamma$ represents the spatial diversity in secondary network, i.e. the number of secondary users that can be packed interference free in $\Omega$. And,

\begin{align}
\label{PII}
P_{\mathbb{II}}^{(m,j)} &= \textrm{Pr}\bigl({\overline{E_{t_x^s}(t_x^p,R_D)}}\mid {\overline{E_{r_x^s}(t_x^p,R_I^p)}},E_{r_x^s}^m(t_x^s,R_I^s)\bigr) \cdot \textrm{Pr}\bigl({\overline{E_{r_x^s}(t_x^p,R_I^p)}}\mid E_{r_x^s}^m(t_x^s,R_I^s)\bigr) \nonumber \\
                        &= P_{\mathbb{IV}}^{(m,j)} \cdot P_{\mathbb{III}}^{(m,j)}
\end{align}

\begin{table}
\caption{Probability Definition}
\label{tab:ProbabilityDefinition}
\centering
\begin{tabular}{|c|c|}
\hline
Notation & Definition\\
\hline
$\gamma$ & $\frac{\left|\mathfrak{C}_{R_I}(.)\right|}{\left|\Omega\right|}$, \\
& Ratio between areas of interference region $\&$ network cell \\
$P_{\mathbb{III}}^{(m,j)}$ & $\textrm{Pr}\bigl({\overline{E_{r_x^s}(t_x^p,R_I^p)}}\mid {E_{r_x^s}^{(m)}(t_x^s,R_I^s)}\bigr)$ \\
$P_{\mathbb{IV}}^{(m,j)}$ & $\textrm{Pr} \bigl({\overline{E_{t_x^s}(t_x^p,R_D)}}\mid{\overline{E_{r_x^s}(t_x^p,R_I^p)}},{E_{r_x^s}^{(m)}(t_x^s,R_I^s)}\bigr)$ \\
\hline
\end{tabular}
\end{table}

After some mathematical manipulation we have

\begin{equation}
\label{probofsuccessfultrans1}
\begin{split}
P_{s \mid (j,K)} &= e^{\biggl(-\lambda \theta_j \Bigl[\pi(R_D^2+R_I^{p^2})-\left|\mathfrak{C}_{R_D}(0,0)\cap\mathfrak{C}_{R_I^p}(R_r^s,0)\right|\Bigr]\biggr)} \\
 & - \sum_{m=1}^{K-1} {K-1\choose m} \gamma^m(1-\gamma)^{K-1-m}P_{\mathbb{III}}^{(m,j)}P_{\mathbb{IV}}^{(m,j)}
\end{split}
\end{equation}
where $P_{\mathbb{III}}^{(m,j)} = \textrm{Pr}\bigl({\overline{E_{r_x^s}(t_x^p,R_I^p)}}\mid {E_{r_x^s}^{(m)}(t_x^s,R_I^s)}\bigr)$ is the probability that there are no primary transmitters within $R_I^p$ radius of a secondary receiver, given that there are $m$ secondary transmitters within $R_I^s$ radius of it. Note that the correlation between the events: ``existence of a secondary transmitter'' and  ``existence of a primary transmitter'' is non-zero because a secondary transmitter transmits if and only if it makes sure that there are no primary transmitters present within radius $R_D$. Likewise $P_{\mathbb{IV}}^{(m,j)} = \textrm{Pr} \bigl({\overline{E_{t_x^s}(t_x^p,R_D)}}\mid{\overline{E_{r_x^s}(t_x^p,R_I^p)}},{E_{r_x^s}^{(m)}(t_x^s,R_I^s)}\bigr)$ is the probability that there are no primary transmitters within $R_D$ radius of a secondary transmitter given that there are $m$ secondary transmitters and no primary transmitter within radius $R_I^s$ and $R_I^p$ of it's receiver, respectively. We obtained $P_{\mathbb{III}}^{(m,j)}$ and $P_{\mathbb{IV}}^{(m,j)}$ approximately as \eqref{PIII} and \eqref{PIV}.

\begin{figure*}
\begin{equation}
\label{PIII}
P_{\mathbb{III}}^{(m,j)} \approx \begin{cases}
\frac{2}{R_I^{s^2}} \int\limits_{0 \leq r \leq R_I^s} r e^{\bigl(-\lambda \theta_j\left|\mathfrak{C}_{R_I^p}(0,0) - \mathfrak{C}_{R_I^p}(0,0)\cap \mathfrak{C}_{R_D}(r,0)\right|\bigr)}dr &\text{$m = 1$} \\
\frac{4}{\pi R_I^{s^4}} \iiint\limits_{\substack{0\leq \varphi \leq \pi \\ 0\leq r_1 \leq R_I^s \\ 0\leq r_2\leq R_I^s}} r_1r_2 e^{\bigl(-\lambda \theta_j\left|\mathfrak{C}_{R_I^p}(0,0) - \mathfrak{C}_{R_I^p}(0,0) \cap \mathfrak{C}_{R_D}(r_1,0) \cup \mathfrak{C}_{R_D}(r_2,\varphi)\right|\big)}d\varphi dr_1dr_2 &\text{$m = 2$}  \\
1 &\text{$m \geq 3$}
\end{cases}
\end{equation}
\end{figure*}

\begin{figure*}
\begin{equation}
\label{PIV}
P_{\mathbb{IV}}^{(m,j)} \approx \begin{cases}
\frac{2}{\pi R_I^{s^2}} \iint\limits_{\substack{0\leq \varphi \leq \pi \\ 0\leq r \leq R_I^s}} r e^{\bigl( -\lambda \theta_j\left|\mathfrak{C}_{R_D}(-R_r^s,0) - \mathfrak{C}_{R_I^p}(0,0)\cup \mathfrak{C}_{R_D}(r,\varphi)\right|\bigr)}d\varphi dr &\text{$m = 1$}  \\
 e^{\bigl(-\lambda \theta_j\left|\mathfrak{C}_{R_D}(-R_r^s+R_I^s/2,0) - \mathfrak{C}_{R_D}(0,0)\right|\bigr)} &\text{$m = 2$} \\
 e^{\bigl( -\lambda \theta_j\left|\mathfrak{C}_{R_D}(-R_r^s+R_I^s,0) - \mathfrak{C}_{R_D}(0,0)\right|\bigr)} &\text{$m \geq 3$}
\end{cases}
\end{equation}
\end{figure*}

\begin{figure}
\centering
\includegraphics[width = 5 in]{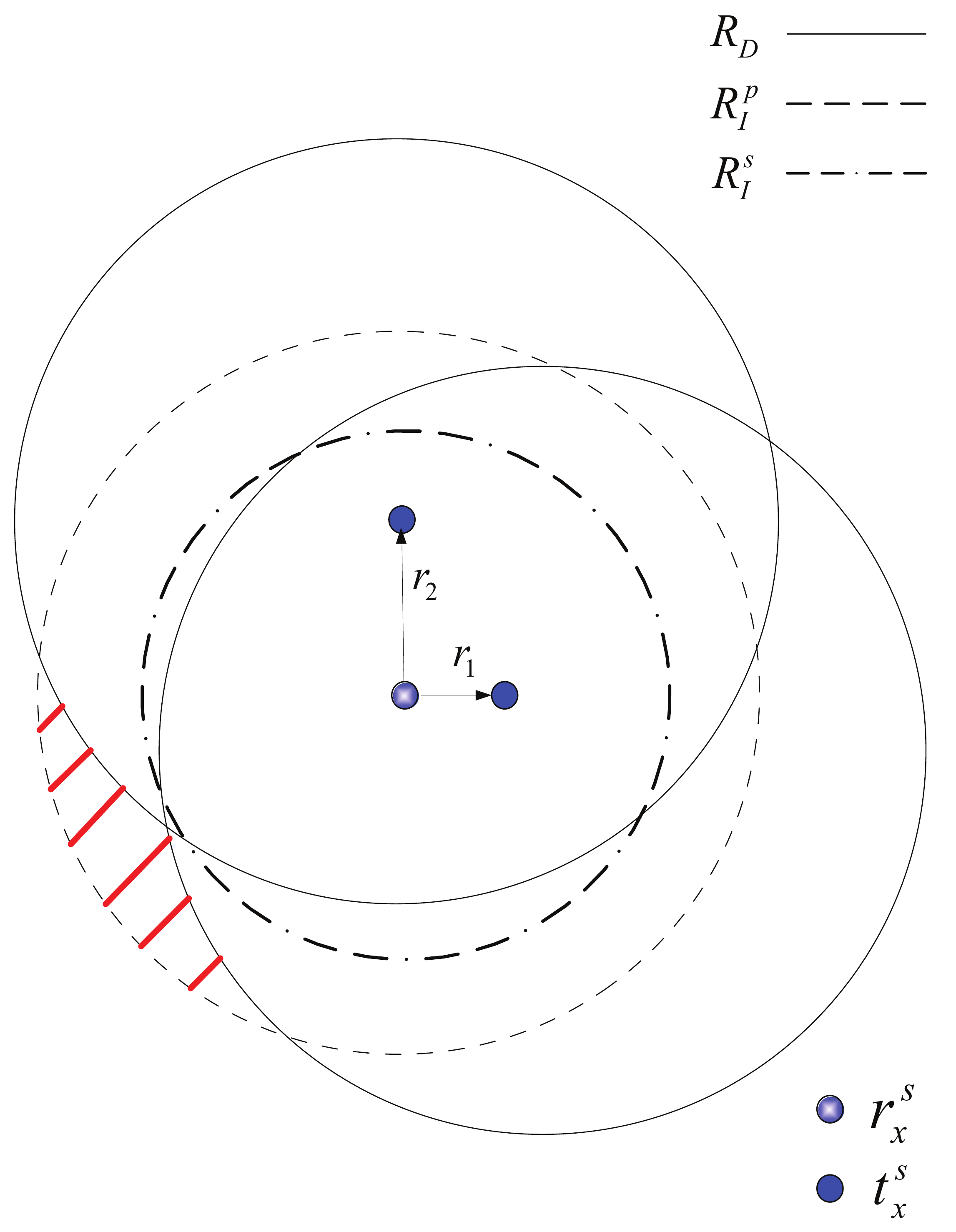}
\caption{Graphical evaluation of $P_{\mathbb{III}}^{(2,j)}$. Solid circle is of radius $R_D$, dashed circle is of radius $R_I^p$ and dashed-dotted circle has radius $R_I^s$. $P_{\mathbb{III}}^{(2,j)}$ equals the probability that there are no primary transmitters in the shaded area.}
\label{fig:PIII}
\end{figure}

Fig. \ref{fig:PIII} illustrates the idea for evaluating $P_{\mathbb{III}}^{(2,j)}$. The two secondary transmitters are assumed to be located at coordinates $[r_1,0]$ and $[r_2,\pi/2]$. The existence of an active secondary transmitter implies that there should not be any primary transmitter present within $R_D$ radius of it. This means that, in Fig. \ref{fig:PIII}, no primary transmitter exists in the union of the two solid circles. Therefore, the probability that there are no primary transmitters within $R_I^p$ radius of the secondary receiver in the presence of the two secondary transmitters reduces to the probability that there are no primary transmitters in the \emph{shaded} area. To obtain $P_{\mathbb{III}}^{(2,j)}$, we need to average this probability over all the possible locations of the two secondary transmitters inside the dashed-dotted circle. For $m > 3$, since $R_D$ is usually greater than $R_I$'s we can assume that the union of the solid circles (the area that secondary detectors encompass) almost cover the whole dashed circle (primary users interference area), resulting in $P_{\mathbb{III}}^{(m,j)}=1$.

\begin{figure}
\centering
\includegraphics[width = 5 in]{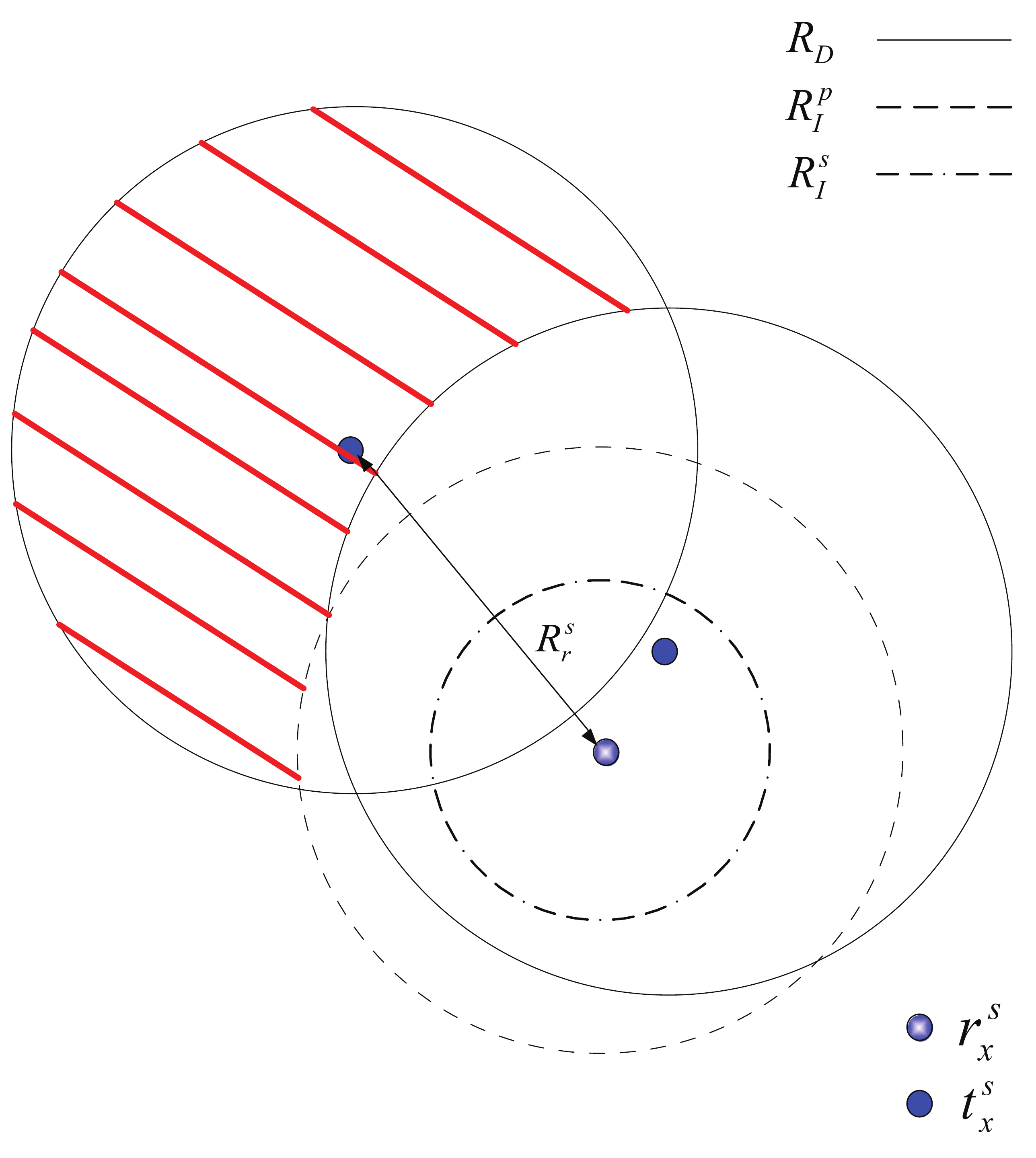}
\caption{Graphical evaluation of $P_{\mathbb{IV}}^{(1,j)}$. Solid circle is of radius $R_D$, dashed circle is of radius $R_I^p$ and dashed-dotted circle has radius $R_I^s$. $P_{\mathbb{IV}}^{(1,j)}$ equals the probability that there are no primary transmitters in the shaded area.}
\label{fig:PIV}
\end{figure}

Similarly, Fig. \ref{fig:PIV} illustrates the idea for evaluating $P_{\mathbb{IV}}^{(1,j)}$. The probability that there are no primary transmitters present within $R_D$ radius of the secondary transmitter, given that there is one secondary transmitter and no primary transmitter within radii $R_I^s$ and $R_I^p$ of its receiver reduces to the probability that there are no primary transmitters in the \emph{shaded} area of Figure \ref{fig:PIV}. Again by averaging over all the positions of the secondary transmitter inside the dashed-dotted circle we obtain the $P_{\mathbb{IV}}^{(1,j)}$. Due to computational complexity, for $m = 2$ and $m \geq 3$, we assume that the net effect of all the secondary transmitters is as if there are two secondary transmitter $R_r^s - R_I^s/2$  and $R_r^s - R_I^s$ distant apart from each other respectively. Therefore $P_{\mathbb{IV}}^{(m,j)}$ approximates as the probability that there are no primary transmitters in $\mathfrak{C}_{R_D}(-R_r^s+R_I^s/2,0) - \mathfrak{C}_{R_D}(0,0)$ and $\mathfrak{C}_{R_D}(-R_r^s+R_I^s,0) - \mathfrak{C}_{R_D}(0,0)$ respectively.

Using the same idea as previous section the throughput of a single secondary user given there are totally $l$ active secondary users can be derived as

\begin{equation}
\label{singlethroughput}
C_{s_1 \mid l} = \sum_{j=1}^{N}\sum_{K=1}^{l}{l\choose K} P_j^K(1-P_j)^{(l-K)} P_{s\mid (j,K)} C_j
\end{equation}
And Finally, the expected {\it secondary network throughput} can be found as:

\begin{equation}
\label{totalthroughput}
C_s = \sum_{l=1}^{M}{M\choose l} q^l(1-q)^{(M-l)} l C_{s_1 \mid l}
\end{equation}

\begin{figure}
\centering
\includegraphics[width=5.5 in]{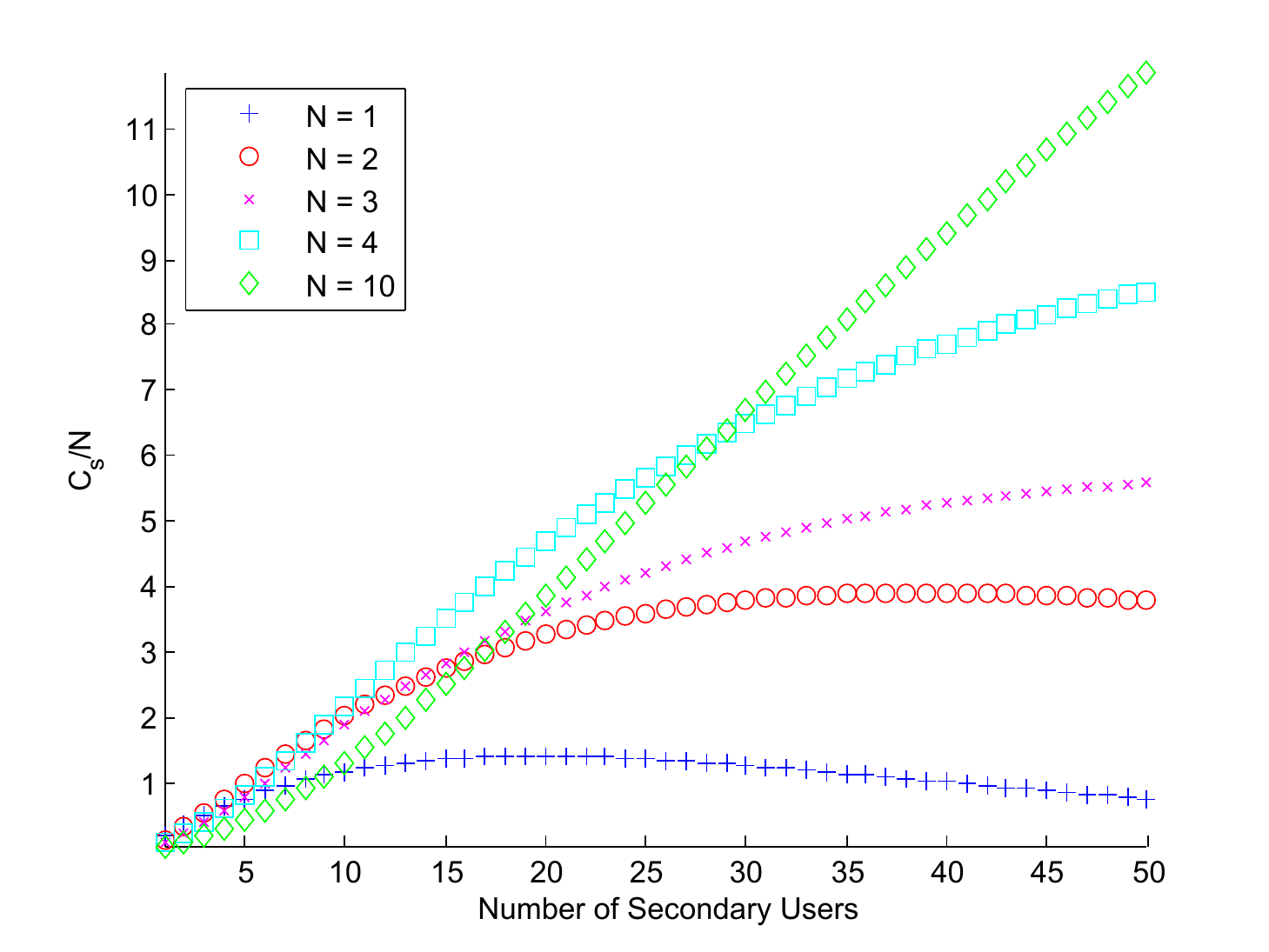}
\caption{The variation of the normalized secondary throughput with respect to the number of the secondary users and the number of frequency channels. The system parameters are set as:$R_I^{s,p}=R_r^{s,p}=1$, $\xi = 0.2$, $q = 0.3$, $\lambda = \frac{1}{1.5^2}$, $\gamma = 0.1$, $\rho = 0.15$}
\label{fig:CsVSM}
\end{figure}

Figure \ref{fig:CsVSM} depicts the normalized secondary user throughput with respect to different values of $M$ and $N$. As the number of secondary users excessively  increases the secondary network throughput drops due to the increase of interference among secondary users. As you can see in this figure, the maximum secondary throughput happens approximately at $M \approx N/(q\gamma)$. As mentioned earlier $1/\gamma$ represents the \emph{spatial} diversity in the secondary network. $N$ can be viewed as the \emph{spectral} diversity for secondary users and $1/q$ represents the average number of the times that a secondary user backs off from transmission, or in other words gives way to other secondary users to capture the spectral opportunity. Therefore, the optimal number of secondary users that can coexist in the network approximately equals the total \emph{spatial} diversity times \emph{spectral} diversity times the back off factor.  

We prove the aforementioned conjecture ($M^* \approx N/(q\gamma)$) to be true analytically in the following special case. To simplify the formulas, assume $R_I = R_r = R$ and all channels have the same capacity and spectrum occupancy statistics, i.e. $C_j = C$ and $\theta_j = \theta$ for $\forall j \in \mathcal{N}$, which amounts to $P_j = 1/N$ for $\forall j \in \mathcal{N}$. Furthermore, assume that $R_D = R_I^s + R_r^p = 2R$ which corresponds to error free detection of the spectral opportunities by secondary users. Thus, we can rewrite \eqref{singlethroughput} as

\begin{equation}
\label{singlethroughputsimplified}
C_{s_1 \mid l} = C N \sum_{K=1}^{l}{l\choose K} (\frac{1}{N})^K(1-\frac{1}{N})^{(l-k)} P_{s\mid K}
\end{equation}
In which we substituted $P_{s\mid (j,K)}$ as $P_{s\mid K}$ because its value is the same across all the channels. In what Follows, we compute each item of $P_{s\mid K}$ in \eqref{probofsuccessfultrans1} separately.

\begin{align}
\label{firstpartaprox}
exp\biggl(-\lambda \theta \Bigl[\pi(R_D^2+R_I^{p^2})-\left|\mathfrak{C}_{R_D}(0,0)\cap\mathfrak{C}_{R_I^p}(R_r^s,0)\right|\Bigr]\biggr) \nonumber \\
= exp\biggl(-\lambda \theta \Bigl[4\pi R^2 - \left|\mathfrak{C}_{2R}(0,0)\cap\mathfrak{C}_{R}(R,0)\right|\Bigr]\biggr) \nonumber \\
= exp\biggl(-\lambda \theta \Bigl[4\pi R^2 - \left|\mathfrak{C}_{R}(R,0)\right|\Bigr]\biggr) \nonumber \\
= e^{- 3 \pi \lambda \theta R^2}
\end{align}

Now we simplify the $P_{\mathbb{III}}^{(m)}$ in \eqref{PIII}. For $m = 1$ we have:

\begin{align}
\label{PIIIm1aprox}
P_{\mathbb{III}}^{(1)} &= \frac{2}{R_I^{s^2}} \int\limits_{0 \leq r \leq R_I^s} r e^{\bigl(-\lambda \theta\left|\mathfrak{C}_{R_I^p}(0,0) - \mathfrak{C}_{R_I^p}(0,0)\cap \mathfrak{C}_{R_D}(r,0)\right|\bigr)}dr \nonumber \\
&= \frac{2}{R^2} \int\limits_{0 \leq r \leq R} r e^{\bigl(-\lambda \theta \left|\mathfrak{C}_{R}(0,0) - \mathfrak{C}_{R}(0,0)\cap \mathfrak{C}_{2R}(r,0)\right|\bigr)}dr
\end{align}
Notice that $\mathfrak{C}_{R}(0,0) \cap \mathfrak{C}_{2R}(r,0)$ equals $\mathfrak{C}_{R}(0,0)$ for $r \in [0,R]$. Therefore, we have $P_{\mathbb{III}}^{(1,j)} = 1$. Similarly, for $m = 2$ we have:

\begin{align}
\label{PIIIm2aprox}
P_{\mathbb{III}}^{(2)} &= \frac{4}{\pi R_I^{s^4}} \iiint\limits_{\substack{0\leq \varphi \leq \pi \\ 0\leq r_1 \leq R_I^s \\ 0\leq r_2\leq R_I^s}} r_1r_2 e^{\bigl(-\lambda \theta_j\left|\mathfrak{C}_{R_I^p}(0,0) - \mathfrak{C}_{R_D}(r_1,0)\cup \mathfrak{C}_{R_D}(r_2,\varphi)\right|\big)}d\varphi dr_1dr_2 \nonumber \\
&= \frac{4}{\pi R^4} \iiint\limits_{\substack{0\leq \varphi \leq \pi \\ 0\leq r_1 \leq R \\ 0\leq r_2\leq R}} r_1r_2 e^{\bigl(-\lambda \theta\left|\mathfrak{C}_{R}(0,0) - \mathfrak{C}_{2R}(r_1,0)\cup \mathfrak{C}_{2R}(r_2,\varphi)\right|\big)}d\varphi dr_1dr_2 \nonumber \\
&= 1
\end{align}
Thus, we have $P_{\mathbb{III}}^{(m)} = 1$ for $\forall m$. Now by assuming $P_{\mathbb{IV}}^{(m)} = e^{- 3 \pi \lambda \theta R^2}$ for simplicity, we get:

\begin{align}
\label{Psk}
P_{s\mid K} &= e^{- 3 \pi \lambda \theta R^2} - \sum_{m=1}^{K-1} {K-1\choose m} \gamma^m(1-\gamma)^{K-1-m}P_{\mathbb{IV}}^{(m,j)} \nonumber \\
   &= e^{- 3 \pi \lambda \theta R^2} (1-\gamma)^{K-1}
\end{align}

By inserting \eqref{Psk} into \eqref{singlethroughputsimplified}, we obtain the conditional throughput of a single secondary user pair:

\begin{align}
\label{singlethroughputaprox}
C_{s_1 \mid l} &= e^{- 3 \pi \lambda \theta R^2} CN \sum_{K=1}^{l}{l\choose K} (\frac{1}{N})^K(1-\frac{1}{N})^{(l-k)} (1-\gamma)^{K-1} \nonumber \\
   &= e^{- 3 \pi \lambda \theta R^2} \frac{CN}{1-\gamma} \sum_{K=1}^{l}{l\choose K} (\frac{1-\gamma}{N})^K(1-\frac{1}{N})^{(l-k)} \nonumber \\
   &= e^{- 3 \pi \lambda \theta R^2} \frac{CN}{1-\gamma} \biggl[(1-\frac{\gamma}{N})^l-(1-\frac{1}{N})^l \biggr]
\end{align}

Finally by substituting \eqref{singlethroughputaprox} into \eqref{totalthroughput}, we obtain the total secondary networks throughput under the simplified system setting:

\begin{align}
\label{totalthroughputaprox}
C_s &= \sum_{l=1}^{M}{M\choose l} q^l(1-q)^{(M-l)} l C_{s_1 \mid l} \nonumber \\
    &= e^{- 3 \pi \lambda \theta R^2} \frac{C}{1-\gamma} N \biggl[ \sum_{l=1}^{M} l {M\choose l} \bigl(q(1-\frac{\gamma}{N})\bigr)^l(1-q)^{(M-l)} - \sum_{l=1}^{M}l {M\choose l} \bigl(q(1-\frac{1}{N})\bigr)^l(1-q)^{(M-l)} \biggr] \nonumber \\
    &= e^{- 3 \pi \lambda \theta R^2} \frac{Cq}{1-\gamma} NM \biggl[ (1-\frac{\gamma}{N})(1-\frac{q\gamma}{N})^{M-1} - (1-\frac{1}{N})(1-\frac{q}{N})^{M-1} \biggr]
\end{align}

First we show that $C_s(M)$ is quasi-concave for $M \geq 1$. Thus, there exists a unique $M$ that maximizes the secondary network throughput. In order to do so, we prove that there exist a $M^*$ (which is also the optimal number of the secondary users) such that $\partial C_s(M)/\partial M > 0 $ for $M < M^*$ and $\partial C_s(M)/\partial M < 0 $ otherwise.

\begin{align}
\label{derivative}
\frac{\partial C_s(M)}{\partial M} &= e^{- 3 \pi \lambda \theta R^2} \frac{Cq}{1-\gamma} N \biggl[(1-\frac{\gamma}{N})(1-\frac{q\gamma}{N})^{M-1}\bigl(1+M\ln(1-\frac{q\gamma}{N})\bigr) \nonumber \\
&- (1-\frac{1}{N})(1-\frac{q}{N})^{M-1}\bigl(1+M\ln(1-\frac{q}{N})\bigr) \biggr]
\end{align}

Note that since $\gamma < 1$, we can neglect the effect of the second term and suppose that

\begin{equation}
\label{derivative2}
\frac{\partial C_s(M)}{\partial M} = e^{- 3 \pi \lambda \theta R^2} \frac{Cq}{1-\gamma} N(1-\frac{\gamma}{N})(1-\frac{q\gamma}{N})^{M-1}\bigl(1+M\ln(1-\frac{q\gamma}{N})\bigr)
\end{equation}

Form \eqref{derivative2}, we can easily derive the aforementioned $M^*$ and the optimal number of the secondary users as

\begin{equation}
\label{aproxoptimalM}
\begin{split}
M^* &= \frac{-1}{\ln(1-\frac{q\gamma}{N})} \\
&\approx \frac{N}{q\gamma}
\end{split}
\end{equation}

\chapter{Randomized Sensing Scheme Based on CSMA/CA Protocol}
\label{ThroughputAnalysisBasedonCSMA/CAProtocol}




In this chapter we derive the optimal cognitive MAC protocol for wireless ad-hoc networks under the assumption that secondary users employ CSMA/CA protocol (\emph{Carrier Sense Multiple Access with Collision Avoidance}) to limit the interference among each other. According to CSMCA/CA protocol, a station wishing to transmit has to first listen to the channel for a random \footnote{We assume it to follow Exponential Distribution.} period of time so as to check for any activity on the channel. If the channel is detected to be ``busy'' the station has to defer its transmission until the subsequent time slot. If the channel is perceived ``idle'', the station sends a ``jamming'' signal telling all other stations not to transmit, and then sends its packet. By doing so it guarantees that one and only one user accesses the shared medium at a time.

The rest of the chapter is organized as follows. We first assume that the secondary users can only sense and access one channel at a time and compare the results with that of Section \ref{ThroughputAnalysisBasedonALOHAProtocol}-\ref{idealscenario}. Then in Section \ref{ThroughputAnalysisBasedonCSMA/CAProtocol}-\ref{MultipleChannelDetection} we extend the analysis to the case in which secondary users can sense and access up to $S \leq N$ frequency channels at a time and show that we can enhance the secondary network throughput considerably by sensing multiple channels. Moreover, we introduce a heuristic sensing scheme that generalizes the sensing scheme proposed in Section \ref{suboptimalsensingscheme} and compare its performance against the optimal sensing strategy. In both previous cases we assume that the secondary users are capable of perfectly detecting the primary signal. We conclude our study by considering the impact of the detection error on the design of the cognitive MAC protocol in Section \ref{ThroughputAnalysisBasedonCSMA/CAProtocol}-\ref{MultipleChannelDetectionwithError}.

We follow the same system assumptions as of Sections \ref{ThroughputAnalysisBasedonALOHAProtocol}-\ref{systemmodel} and \ref{ThroughputAnalysisBasedonALOHAProtocol}-\ref{idealscenario} unless otherwise mentioned. In the next section we briefly recapitulate the system model.

\section{System Model}
\label{systemmodel2}

Consider a frequency spectrum consisting of $N$ independent channels each with bandwidths $W_j$, $j \in \mathcal{N}$. Each channel is licensed to a time-slotted primary user. The spectrum occupancy statistics of channels are independent of each other and follow a Bernoulli distribution with probability $\theta_j$, $j \in \mathcal{N}$ of being active for each channel.

Channels are also available to $M$ secondary users who are seeking opportunities to access the spectrum. The secondary user network is also time-slotted. The spectrum occupancy statistics of primary channels are available to all secondary users.

Assuming a fixed physical layer scheme in all channels, the rate at which users communicate is proportional to their available bandwidth, i.e.

\begin{equation}
\nonumber
\label{proportionality2}
C_j = a \cdot W_j
\end{equation}
where  the proportionality constant $a$ depends on the physical layer signaling scheme.

At the beginning of each time slot each secondary user decides on which channel(s) to sense according to the scheme introduced in \eqref{defsensscheme}. Based on the sensing outcome, if a particular channel/group is found to be idle, the secondary user initiates the transmission \footnote{In order to compensate for hidden terminal problem, RTS and RTS packets can also be implemented.} after confirming that the channel is still idle (i.e. no other secondary user is using the channel) by monitoring it for a random period of time. Otherwise, it will wait until the next time slot and repeat the same procedure again.

\section{Single Channel Detection}
\label{singlechanneldetection}

In this section we consider the scenario in which $M$ secondary users seek to access the spectrum opportunistically and they are capable of sensing and accessing only a single channel. Secondary users are also equipped with error free detector. As mentioned earlier secondary users employ the \emph{Carrier Sense Multiple Access/Collision Avoidance protocol} to access the idle perceived channels (i.e. after sensing a channel idle, each user waits a random time and if, yet again, senses the channel idle, initiates the transmission). If we assume exponential distribution for the waiting time, the probability that a particular user succeeds in seizing an idle channel among $l$ users can be derived as $\frac{1}{l}$. Therefore it is clear that choosing the channel with largest probability of being idle is no longer optimal. Intuitively, we'd like secondary users sensing strategy to span more spectrum channels with high probability of being idle.

In order to achieve this goal assume each secondary user chooses channel $j$ with probability $P_j$ to sense \footnote{Considering the same $P_j$ for all secondary users does not affect the generality of the scheme.}. Also note that since secondary users utilize perfect spectrum detectors, no collision will occur between primary and secondary users. We can find the optimal sensing strategy that maximizes the total secondary network throughput as follows.

Using CSMA/CA protocol, secondary network will make use of an spectrum opportunity (an idle channel)  as long as that channel is sensed by at leased one secondary user. Therefore, instead of maximizing the secondary network throughput we can minimize the potential throughput that was not utilized by both primary and secondary network. This event happens when a particular channel is not used by primary users (with probability $\overline{\theta_j}$) and has not been sensed by any of the secondary users (with probability $(1-P_j)^M$). Hence, the solution to the following optimization problem characterizes the optimal sensing scheme;

\begin{equation}
\label{singlechanneldetectionlostthroughput}
\begin{split}
\textrm{min} ~~~~&\sum_{j=1}^N C_j\overline{\theta_j}(1-P_j)^M \\
\textrm{s.t} ~~~~&\sum_{j=1}^N P_j = 1 \\
& P_j \geq 0, ~~~\forall j \in \{1 \cdots N\}
\end{split}
\end{equation}
The objective function is clearly convex for $0 \leq P_j \leq 1$ and the constraints are linear inequalities which ensues a convex set. Therefore, we have a convex optimization problem with the corresponding KKT conditions as follow:

\begin{equation}
\begin{split}
p_j^*(\nu^* - M C_j\overline{\theta_j}(1-P_j)^{M-1}) &= 0, ~~~\forall j \\
 M C_j\overline{\theta_j}(1-P_j)^{M-1}& \leq \nu^*, ~~\forall j\\
\sum_{j=1}^N P_j^* &= 1 \\
P_j^* &\geq 0, ~~~\forall j \in \{1 \cdots N\}
\end{split}
\end{equation}
where $\nu$ is a Lagrange multiplier.

For $M \geq 2$ we can obtain the close form optimal solution as:

\begin{equation}
p_j^* = \biggl(1 - \bigl(\frac{\nu}{MC_j\overline{\theta_j}}\bigr)^{1/(M-1)}\biggr)^+ ~~~ j \in \{1 \cdots N\}
\end{equation}
for $\theta_j < 1$ and $p_j^* = 0$ for $\theta_j = 1$. $\nu$ is chosen such that $\sum P_j^* = 1$ is satisfied.

\begin{figure}
\centering
\includegraphics[width=5.5in]{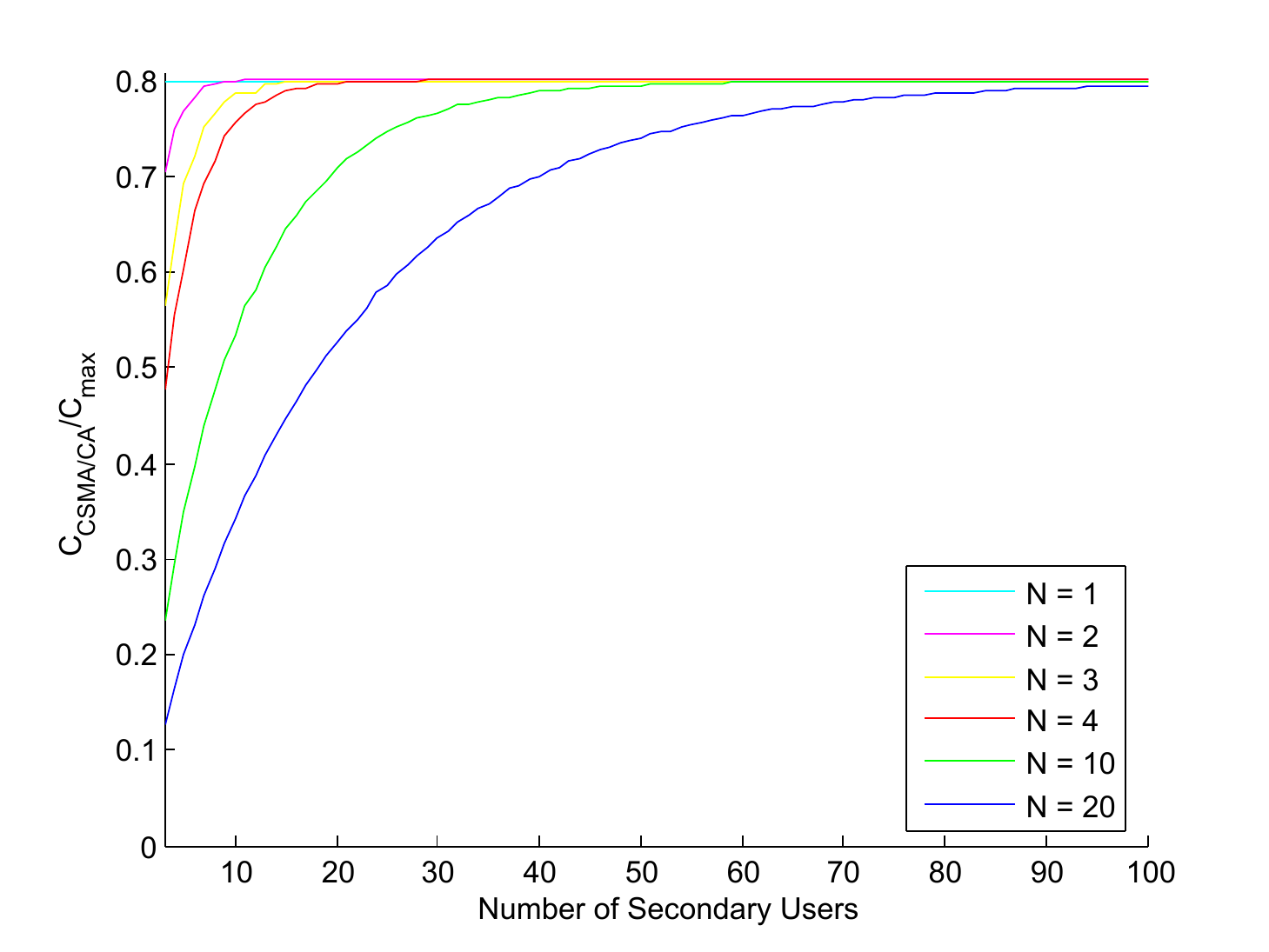}
\caption{Normalized secondary network throughput variation with respect to the number of the secondary users and the number primary users using CSMA/CA protocol. The channel utilization statistic is chosen as $\rho = 0.8$ which corresponds to \emph{highly utilized spectrum}}
\label{fig:optimalgoodputcsmaca}
\end{figure}

Figure \ref{fig:optimalgoodputcsmaca} depicts the normalized optimal secondary network throughput with respect to the number of the secondary and the primary users using CSMA/CA protocol. It is shown that as the number of secondary users tends to infinity, the secondary network can fully utilize the residual capacity, which coincides with  \eqref{singlechanneldetectionlostthroughput}. As the number of frequency channels increases, more secondary users are needed to exhaustively scan all the channels for potential opportunities and utilize the residual capacity to the fullest extend.

\begin{figure}
\centering
\includegraphics[width=6in]{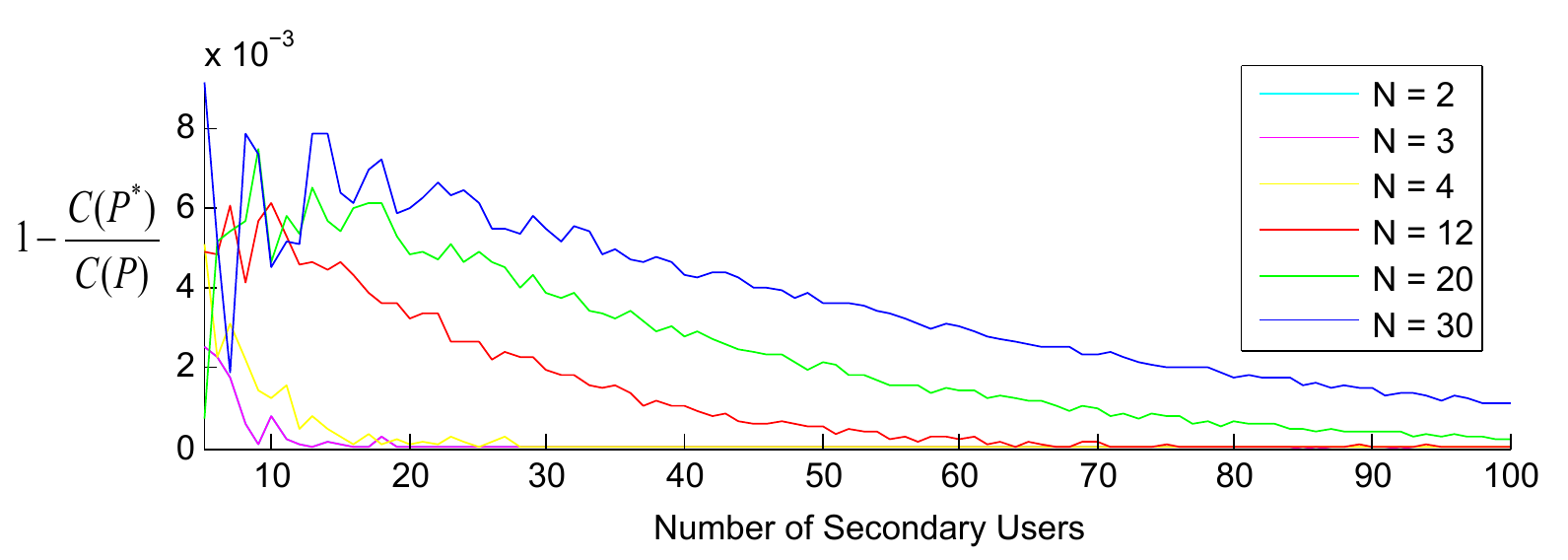}
\caption{Throughput percentage loss caused by using the heuristic sensing scheme instead of the optimal scheme under highly utilized spectrum regime ($\rho = 0.8$).}
\label{fig:optimalVSheuristic}
\end{figure}

In Figure \ref{fig:optimalVSheuristic} we compare the performance of the system under optimal and heuristic sensing scheme introduced in Section \ref{ThroughputAnalysisBasedonALOHAProtocol}-\ref{suboptimalsensingscheme}. For the sake of illustration we denote the secondary network throughput using optimal sensing scheme by $C(P^*)$ and the throughput using heuristic sensing scheme by $C(P)$. Fig. \ref{fig:optimalVSheuristic} demonstrates the throughput loss percentage brought about by using the heuristic sensing scheme instead of the optimal scheme. The plot is drawn in the highly utilized spectrum regime ($\rho = 0.8$). As shown in this figure, the loss percentage is an increasing function of $N$ and decreasing function of $M$ and in all cases negligible. Therefore, we can conclude that there is  almost no advantage in finding the optimal sensing scheme when we can only sense and access a single channel and the heuristic approach works as fine.

\begin{figure}
\centering
\includegraphics[width=6in]{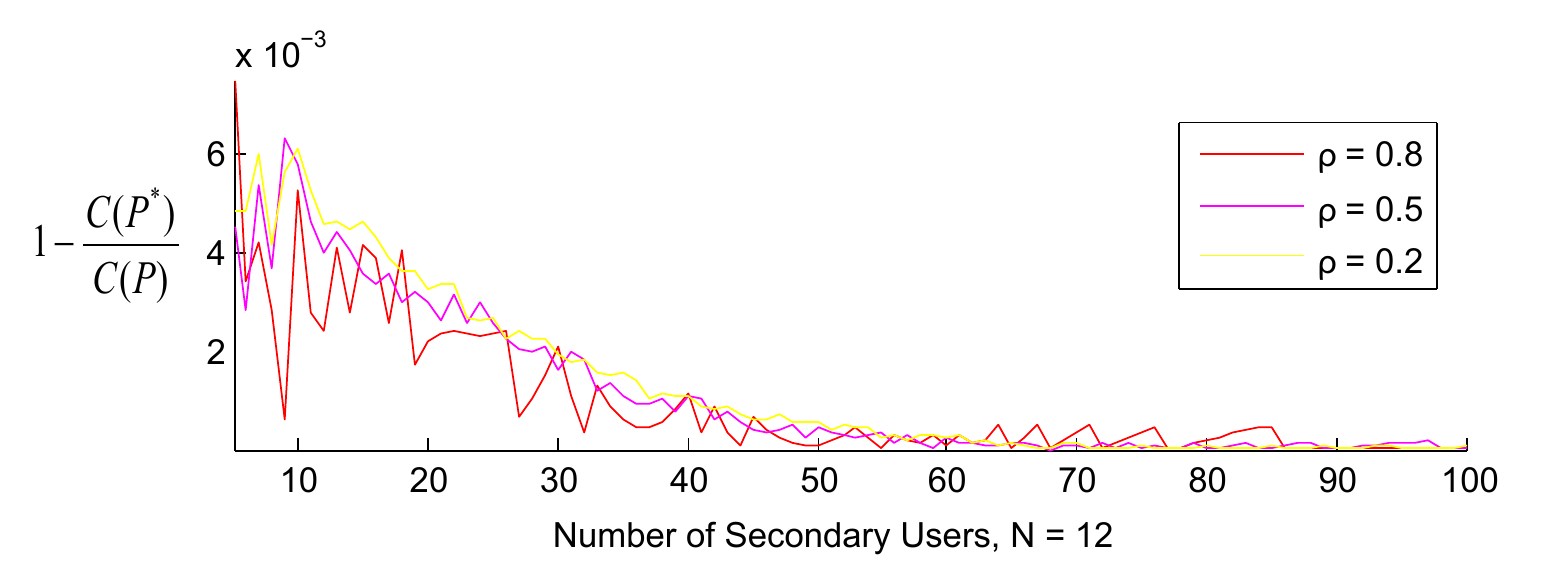}
\caption{Throughput percentage loss caused by using the heuristic sensing scheme instead of the optimal scheme for three channel utilization statistics $\rho = 0.2$ (Rarely Utilized), $\rho = 0.5$ (Fairly Utilized), $\rho = 0.8$ (Highly Utilized) and $N=12$.}
\label{fig:optimalVSheuristicRHO}
\end{figure}

In Figure \ref{fig:optimalVSheuristicRHO} we compare the performance of the system under optimal and heuristic sensing scheme for different channel utilization statistics. As suggested by this figure the performance difference is unaffected by the utilization statistics of the frequency channels.

\section{Multiple Channel Detection}
\label{MultipleChannelDetection}

In this section we consider the scenario in which each secondary user is capable of sensing and accessing multiple ($S \leq N$) channels at a time and analyze how much leverage we can gain by doing so compared to the single channel case. Following the definition in Section \ref{ThroughputAnalysisBasedonALOHAProtocol}-\ref{systemmodel}, let $G_j$ represent the $j$th set (group) of channels that a user can sense and let $\mathcal{G} := \{G_1, \ldots , G_{\kappa}\}$ be the collection of all such groups. In the most general case $\kappa$ equals ${N\choose S}$. From here on we use the words ``group'' and ``channel group'' interchangeably. Without loss of generality we assume that all the secondary users adopt the same assignment for the probability $P_j$ of choosing group $j$ to sense. Secondary users abide by the same access protocol as described in Section \ref{ThroughputAnalysisBasedonCSMA/CAProtocol}-\ref{singlechanneldetection}.

Following the argument of the previous section, our problem reduces to finding the optimal group sensing strategy that minimizes the opportunities that are overlooked by secondary users (note that by doing so we do not expose the primary network to any loss due to the perfect secondary detectors). In this case, the probability that a particular channel (e.g. $i$) is not sensed by a secondary user is $\sum_{j:i\notin G_j} P_j$. consequently we can formulate the overlooked capacity as

\begin{equation}
\label{Cunusedmult}
C_{\textrm{unutilized}} = \sum_{i=1}^N C_i\overline{\theta_i} \bigl(1-\sum_{j:i\in G_j} P_j\bigr)^M
\end{equation}

and the optimal sensing strategy can be obtained by solving the optimization problem:

\begin{equation}
\begin{split}
\textrm{min} ~~~~&\sum_{i=1}^N C_i\overline{\theta_i} \bigl(1-\sum_{j:i\in G_j} P_j\bigr)^M \\
\textrm{s.t} ~~~~&\sum_{j=1}^{\kappa} P_j = 1 \\
& P_j \geq 0, ~~~\forall j \in \{1,\cdots,\kappa\}
\end{split}
\end{equation}
with the following KKT conditions

\begin{equation}
\begin{split}
p_j^*(\nu^* - M \sum_{i=1}^N C_i\overline{\theta_i} \bigl(1-\sum_{j:i\in G_j} P_j\bigr)^{M-1} &= 0 \\
M \sum_{i=1}^{\kappa} C_i\overline{\theta_i} \bigl(1-\sum_{j:i\in G_j} P_j\bigr)^{M-1} &\leq \nu^* \\
\sum_{j=1}^N P_j^* &= 1 \\
P_j^* &\geq 0, ~~~\forall j \in \{1,\cdots,\kappa\}
\end{split}
\end{equation}

\begin{figure}
\centering
\includegraphics[width=6in]{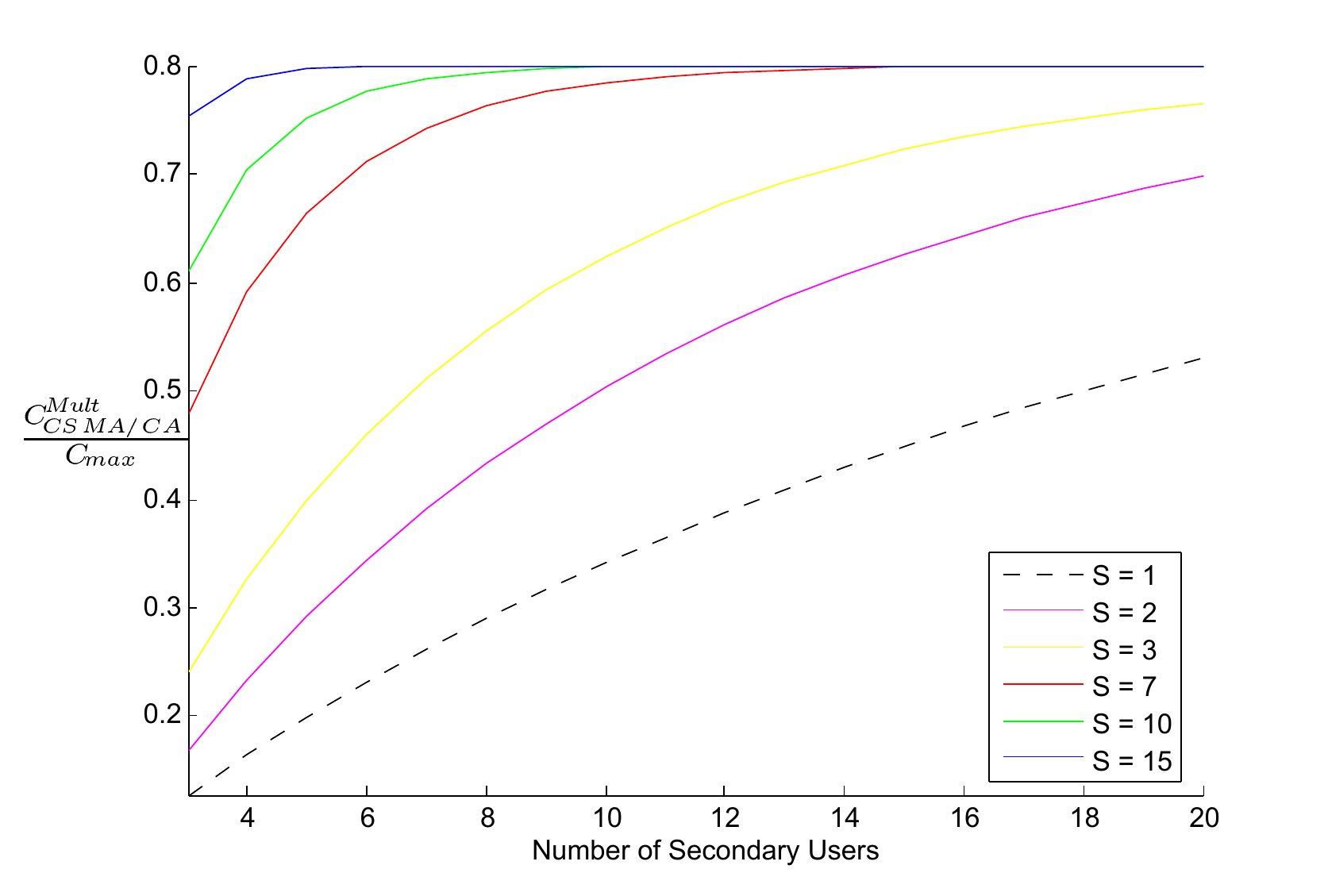}
\caption{Normalized throughput of secondary users with multiple channel sensing and accessing capability for channel utilization statistic $\rho = 0.2$ (Rarely Utilized), and $N=20$.}
\label{fig:CMUltS}
\end{figure}
where $\nu$ is a Lagrange multiplier. Figure \ref{fig:CMUltS} illustrates the performance enhancement due to multiple channel sensing and accessing capability. From this figure it is apparent that secondary network can fully utilize the un-used spectrum with less members as the capability of the secondary users to sense and access multiple channels increases. However, it is not always the case that a system comprised of $M$ secondary users with capability of sensing and accessing $S$ channels achieves more throughput than a system with $MS$ secondary users which are capable of sensing and accessing only a single channel at a time. For example, as shown in Figure \ref{fig:CMUltS},  the sum throughput of the network with $6$ secondary users that are capable of sensing and accessing two channel at a time is less than the sum throughput of the network with $12$ secondary users capable of sensing and accessing a single channels at a time.

In the previous section we observed that the heuristic sensing scheme \eqref{searchprobability} achieves almost the same throughput as the optimal scheme. Now we adopt the idea in definition of \eqref{searchprobability} and extend it to a sensing scheme \eqref{multheurscheme} which assigns a probability $P_j$ for sensing group $j$ proportional to the expected residual throughput of that group:

\begin{equation}
\label{multheurscheme}
P_j := \frac{1}{S}\frac{\sum_{i \in G_j}\overline{\theta_i}\cdot C_i}{\sum_{i=1}^N \overline{\theta_i}\cdot C_i}
\end{equation}

\begin{figure}
\centering
\includegraphics[width=6.5in]{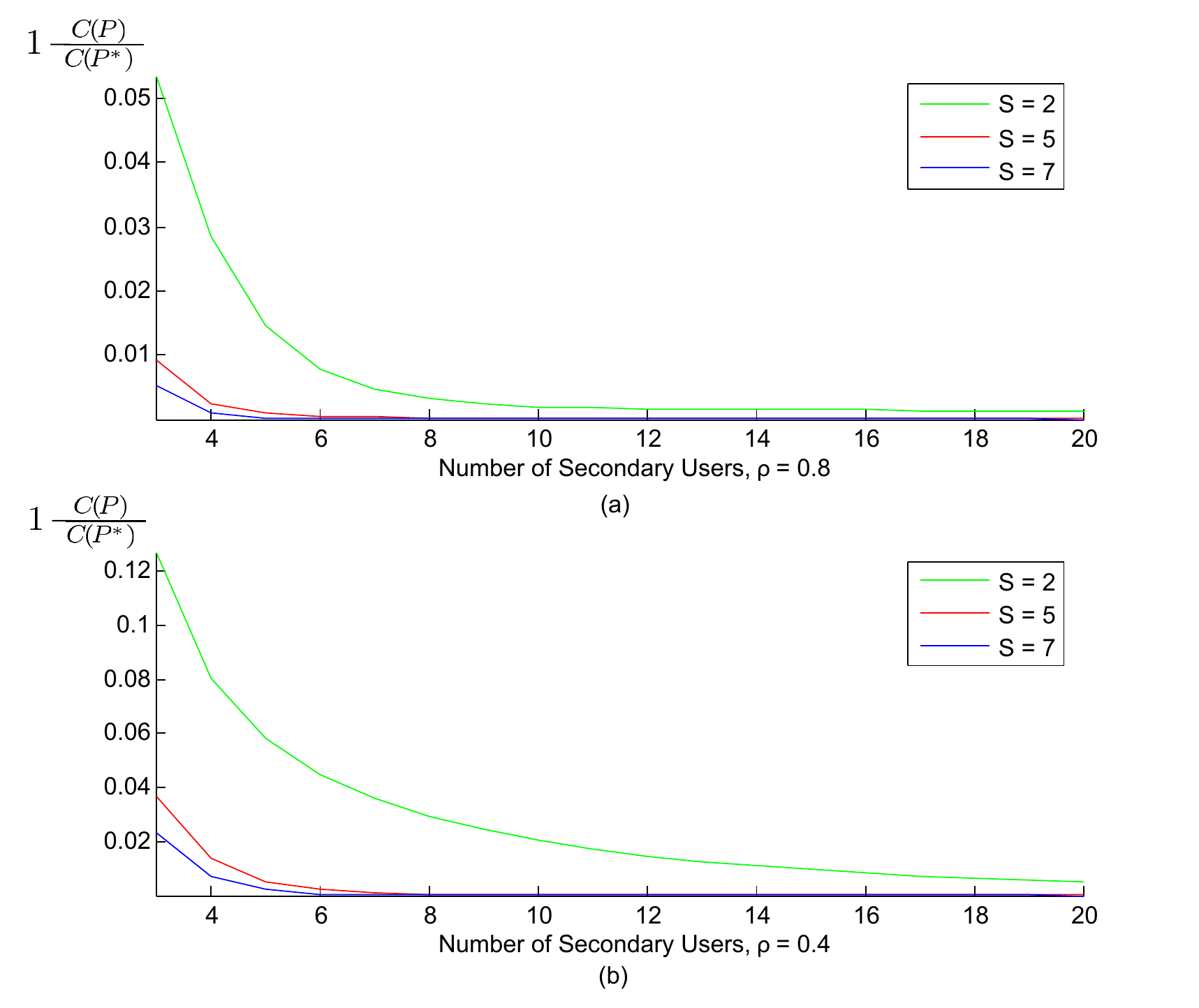}
\caption{Performance comparison between optimal and heuristic sensing scheme when secondary secondary detectors are capable of sensing $S = 2$, $S = 5$ and $s = 7$ at a time. The number of primary channels equal $N = 10$. (a) The channel utilization statistic $\rho = 0.8$ (b) The channel utilization statistic $\rho = 0.4$.}
\label{fig:CMUltVSheu}
\end{figure}

Figure \ref{fig:CMUltVSheu} demonstrates the throughput loss percentage introduced by using the heuristic sensing scheme \eqref{multheurscheme} rather than the optimal scheme. Figure \ref{fig:CMUltVSheu} (a) is drawn for the highly utilized spectrum regime ($\rho = 0.8$), various group size ($S$) and $N = 10$. As shown in the plots, the optimal sensing scheme outperforms the heuristic scheme for small values of $S$. But as the secondary users become capable of sensing wider bandwidths, the throughput loss percentage decreases. Figure \ref{fig:CMUltVSheu} (b) suggests that  the throughput loss percentage grows bigger as the Primary network utilization efficiency increases (i.e. the frequency spectrum becomes more congested).

\section{Multiple Channel Detection with Error}
\label{MultipleChannelDetectionwithError}

This section extends the results in Section \ref{ThroughputAnalysisBasedonCSMA/CAProtocol}-\ref{MultipleChannelDetection} to the scenario in which  secondary users detection outcome is prone to errors. Therefore, unlike previous sections, it is likely that secondary users interfere with the primary users communication. Consequently, the sensing and access scheme should be modified to compensate for the  possible collision caused by erroneous detection results. Now we define the secondary users \emph{detector} related parameter. Let the events $H_0$ and $H_1$ represent:

\begin{equation}
\begin{split}
H_0 &: \textrm{The event that a frequency band is {\it idle}} \\
H_1 &: \textrm{The event that a frequency band is {\it busy}.}
\end{split}
\end{equation}

We also define the Receiver Operation Characteristic (ROC) for secondary detectors as

\begin{equation}
\begin{split}
\alpha &= \textrm{Pr$\bigl($detect $H_0 \mid H_1$ is true$\bigr)$} \\
1- \beta & = \textrm{Pr$\bigl($detect $H_1 \mid H_0$ is true$\bigr)$}
\end{split}
\end{equation}
In which $\alpha$ represents the probability of {\it miss-detection} and $1-\beta$ represents the probability of {\it false-alarm}.

Due to detection errors when the channel is detected idle there is a chance that secondary users will cause interference to primary users. Also when a channel is detected busy there is a chance that we are overlooking an opportunity. Since secondary users channel access scheme can only depend on the detectors outcome, we introduce the following two parameters to make up for these cases. Let $\phi$ be the random variable that indicates whether the channel is being access by a secondary user with $\phi = 1$ denoting the secondary user accessing the channel. We assume that the secondary users accesses the channel with probability $f^0$ if the channel/group is detected idle and $f^1$ if it is detected busy;

\begin{equation}
\begin{split}
f^0 &:= \textrm{Pr}(\phi=1 \mid \textrm{detect $H_0$} )  \\
f^1 &:= \textrm{Pr}(\phi=1 \mid \textrm{detect $H_1$} )
\end{split}
\end{equation}

Primary network collision constraint bounds the secondary users to limit their probability of collision in each channel to at most $\xi$. In order to meet the primary network collision constraint easier and further simplify the formulations, we assume $f^1 = 0$. By doing so, the secondary users do not transmit when the channel is detected busy (although it's likely that the secondary network may loose some throughput).

\begin{table}
\caption{Detector Parameters}
\centering
\begin{tabular}{|c|c|}
\hline
Parameter & Definition\\
\hline
$H_0$ & Channel is idle \\
$H_1$ & Channel is busy \\
$\alpha$ & Probability of miss-detection\\
$1 - \beta$ & Probability of false-alarm\\
$\phi$ & Channel access indicator\\
$f^0$ & Probability of accessing the channel while perceived idle\\
$f^1$ & Probability of accessing the channel while perceived busy\\
\hline
\end{tabular}
\label{tab:DetectorParameters}
\end{table}

We'd like to find the optimal sensing (i.e $P$) and access (i.e. $f^0$) policies which minimize the overlooked throughput by secondary network and yet satisfy the collision constraint. Following the same approach as previous sections we identify following three cases where an opportunity is overlooked: (assume channel $i$ is idle)

Case $1$: none of the secondary users sense channel $i$ and it is idle. (Lost throughput $C_\mathbb{I}$)

Case $2$: Some secondary users sense channel $i$, it is idle but they all detect it busy, others don't sense it. (Lost throughput $C_\mathbb{II}$)

Case $3$: Some secondary users sense channel $i$, it is idle, they detect it idle, but they do not access it (with probability $1-f^0$), others don't sense it. (Lost throughput $C_\mathbb{III}$)

We then determine the total {\it lost} throughput for each case above as:

\begin{eqnarray}
\nonumber C_\mathbb{I} + C_\mathbb{II} &=& \sum_{i = 1}^N \biggr[ C_i\overline{\theta_i} \sum_{k=0}^{M} {M \choose k}
\big[ (\sum_{j:i \in G_{j}} p_{j})(1-\beta) \big]^{k}\\
\nonumber &.& (1-\sum_{j:i \in G_{j}} p_{j})^{M-k} \biggl]\\
\nonumber &=& \sum_{i = 1}^N \biggr[ C_i\overline{\theta_i} \big[ (1-\beta)(\sum_{j:i \in G_{j}} p_{j}) + (1-\sum_{j:i \in G_{j}} p_{j})\big]^{M}\biggl] \\
          &=& \sum_{i = 1}^N \biggr[ C_i\overline{\theta_i} \big( 1- \beta \sum_{j:i \in G_{j}} p_{j} \big)^{M}\biggl]
\end{eqnarray}


\begin{eqnarray}
\label{CIII}
\nonumber C_\mathbb{III } &=& \sum_{i = 1}^N \biggr[ C_i\overline{\theta_i} \sum_{k=0}^{M} {M \choose k}
\big[ (\sum_{j:i \in G_{j}} p_{j}) \beta (1-f^0) \big]^{k}(1-\sum
p_{j})^{M-k} \biggl] \\
\nonumber &=& \sum_{i = 1}^N \biggr[ C_i\overline{\theta_i} \big[ \beta(\sum_{j:i \in G_{j}} p_{j})(1-f^0) + (1-\sum_{j:i \in G_{j}} p_{j})\big]^{M} \biggl] \\
&=& \sum_{i = 1}^N \biggr[ C_i\overline{\theta_i} \big[ 1- \big(1-\beta(1-f^0)\big)\sum_{j:i \in G_{j}} p_{j} \big]^{M} \biggl]
\end{eqnarray}

The \emph{objective function} for our minimization problem would be $Z = C_\mathbb{I} + C_\mathbb{II} + C_\mathbb{III}$. According to the previous sections, $C_\mathbb{I} + C_\mathbb{II}$ is convex. Now we only need to analyze the convexity of $Z_i = \bigl[1- \big(1-\beta(1-f^0)\big)\sum_{j:i \in G_{j}} p_{j} \bigr]^{M}$. Consider the Hessian matrix of $Z$;

\begin{equation}
\nabla Z_i = \left[
\begin{array}{cc}-MZ_i^{M-1}\beta(\sum_{j:i \in G_{j}} p_{j}) &
-MZ_i^{M-1}(1-\beta(1-f^0))[1 \cdots 1]
\end{array}\right]
\end{equation}

\begin{equation}
\label{hessian}
\nabla^{2} Z_i = \left[
\begin{array}{cccc}
a & b & \cdots & b\\
b & c & \cdots & c\\
\vdots & \vdots & \ddots & \vdots \\
b & c & \cdots & c
\end{array}
\right]
\end{equation}
where

\begin{subequations}
\label{abc}
\begin{align}
a &= M(M-1)\beta^{2}(\sum_{j:i \in G_{j}} p_{j})^{2}Z_i^{M-2} \\
b &= M\beta Z_i^{M-2}\big((M-1)(\sum_{j:i \in G_{j}} p_{j})(1-\beta(1-f^0)) - Z_i \big) \\
c &= M(M-1)\bigl(1-\beta(1-f^0)\bigr)^{2}Z_i^{M-2}
\end{align}
\end{subequations}

$C_\mathbb{III}$ is convex if and only if the Hessian matrix $\nabla^{2} Z_i$ is positive semi-definite. Lets consider the Schur complement of the block $a$ of the matrix $\nabla^{2} Z_i$ (provided that $a > 0$):

\begin{align}
\label{schurcomplement}
SC(\nabla^{2} Z_i ; a) &= \left[
\begin{array}{ccc}
c & \cdots & c\\
\vdots & \ddots & \vdots \\
c & \cdots & c
\end{array}
\right]
- \frac{1}{a} \left[
\begin{array}{c}
b \\
\vdots \\
b\end{array}
\right]
\left[
\begin{array}{ccc}
b \cdots b
\end{array}
\right] \nonumber \\
&= \bigl(c - \frac{b^2}{a}\bigr) \left[
\begin{array}{ccc}
1 & \cdots & 1\\
\vdots & \ddots & \vdots \\
1 & \cdots & 1
\end{array}
\right]
\end{align}

According to the properties of the Schur complement, $\nabla^{2} Z_i$ is positive semi-definite if and only if $a > 0$ and $SC(\nabla^{2} Z_i ; a)$ is positive semi-definite. Now we check each of this conditions separately.

Condition $1$: \hspace{6 mm} $a > 0$ \hspace{1 mm} $\Rightarrow$

\begin{align}
\label{convexitycondition1}
&(1-\beta(1-f^0))\sum_{j:i \in G_{j}} p_{j} < 1 \qquad \Rightarrow \nonumber \\
& \sum_{j:i \in G_{j}} p_{j} \leq 1 < \frac{1}{1-\beta(1-f^0)} \qquad \forall i \in \mathcal{N}
\end{align}
which always holds given $M \geq 2$, $\beta > 0$ and $f^0 > 0$.

Condition $2$: \hspace{6 mm} $\bigl(c - \frac{b^2}{a}\bigr) \geq 0$ \hspace{1 mm} $\Rightarrow$

\begin{equation}
\label{convexitycondition2}
\nonumber
\begin{split}
\bigl[1- \big(1-\beta(1-f^0)\big)&\sum_{j:i \in G_{j}} p_{j} \bigr]^{M} \\
&\leq 2(M-1)(1-\beta(1-f^0))\sum_{j:i \in G_{j}} p_{j}, \qquad \forall i \in \mathcal{N}
\end{split}
\end{equation}

This condition is not convex over $(P,f^0)$ in general. Therefore, in order to guarantee the convexity of \eqref{CIII}, we will satisfy a stricter yet convex inequality:

\begin{align}
\label{convexitycondition3}
\bigl[1- \big(1-\beta\big)&\sum_{j:i \in G_{j}} p_{j} \bigr]^{M} \nonumber \\
&\leq 2(M-1)(1-\beta)\sum_{j:i \in G_{j}} p_{j}, \qquad \forall i \in \mathcal{N}
\end{align}

Note that for large enough $M$ condition \eqref{convexitycondition3} almost always holds:

\begin{align}
\bigl[1- \big(1-\beta\big)\sum_{j:i \in G_{j}} p_{j} \bigr]^{M} &\leq \bigl[1- \big(1-\beta\big) \sum_{j:i \in G_{j}} p_{j} \bigr] \nonumber \\
&\leq 2(M-1)(1-\beta)\sum_{j:i \in G_{j}} p_{j}, \qquad \forall i \in \mathcal{N} \nonumber \\
\Rightarrow \sum_{j:i \in G_{j}} p_{j} &\geq \frac{1}{(2M-1)(1-\beta)} \qquad \qquad ~~\quad \forall i \in \mathcal{N}
\end{align}

In other words, for large enough $M$, the objective function is convex {\it almost} everywhere on its domain.

As mentioned before, due to detection errors, if we do not limit our access policy we would cause too much interference to the primary users. Next we introduce the constraint associated with the primary network collision constraint. Let $\xi$ be the probability of collision that is tolerable by primary users in each channel. A collision occurs when channel $i$ is busy but it is sensed idle and accessed by at least one secondary user. Therefore, the probability of collision in channel $i$ can be expressed as:

\begin{eqnarray}
\label{ii}
\nonumber P_c^i &=& \theta_{i}\sum_{k=1}^{M} {M \choose k} \big[ (\sum_{j:i \in G_{j}} p_{j})\alpha f^0 \big]^{k} (1-\sum_{j:i \in G_{j}} p_{j})^{M-k}\\
&=& \theta_{i}\bigl[\alpha f^0 \sum_{j:i \in G_{j}} p_{j} + (1-\sum_{j:i \in G_{j}} p_{j})\bigr]^{M} - (1-\sum_{j:i \in G_{j}} p_{j})^{M}
\leq \xi
\end{eqnarray}
for $i \in \mathcal{N}$. We can further restrict this constraint as:

\begin{equation}
\label{quasiconveprime1}
\theta_{i}(1-(1-\alpha f^0)\sum_{j:i \in G_{j}} p_{j})^{M} \leq \xi \qquad \forall i \in \mathcal{N}
\end{equation}

The condition \eqref{quasiconveprime1} is not convex over $(P,f^0)$ in general. Instead, we require the secondary users to satisfy the following convex constraint:

\begin{equation}
\label{quasiconveprime2}
\theta_{i}(1-\sum_{j:i \in G_{j}} p_{j}+ \alpha f^0)^{M} \leq \xi \qquad \forall i \in \mathcal{N}
\end{equation}

Note that \eqref{quasiconveprime2} is a more restricted version of the \eqref{quasiconveprime1}. Thus, by satisfying \eqref{quasiconveprime2} we would meet the primary network collision constraint. This constraint itself is convex and for small $\alpha$ the \eqref{ii} is convex {\it almost} everywhere. We summarize the problem of finding the optimal sensing (i.e $P$) and access (i.e. $f^0$) policies as the following convex minimization problem:

\begin{eqnarray}
\label{finalmulterr}
\begin{split}
\textrm{min} \quad &\sum_{i = 1}^N C_i\overline{\theta_i} \biggr[\big( 1- \beta \sum_{j:i \in G_{j}} p_{j} \big)^{M} + \big(1-(1-\beta(1-f^0))\sum_{j:i \in G_{j}} p_{j}\big)^{M}\biggl] \\
\textrm{s.t}  \quad &\sum_{j=1}^{\kappa} P_j = 1 \\
& \bigl[1- \big(1-\beta\big)\sum_{j:i \in G_{j}} p_{j} \bigr]^{M} \leq 2(M-1)(1-\beta)\sum_{j:i \in G_{j}} p_{j} \qquad i \in \mathcal{N} \\
& \theta_{i}(1-\sum_{j:i \in G_{j}} p_{j}+ \alpha f^0)^{M} \leq \xi \qquad \qquad \qquad \qquad \qquad \quad ~~ i \in \mathcal{N} \\
& P_j \geq 0 \qquad \qquad \qquad \qquad \qquad \qquad \qquad \qquad \quad \qquad \quad j \in \{1, \cdots, \kappa\} \\
& 0 < f^0 \leq 1
\end{split}
\end{eqnarray}

\begin{figure}
\centering
\includegraphics[width=5.5in]{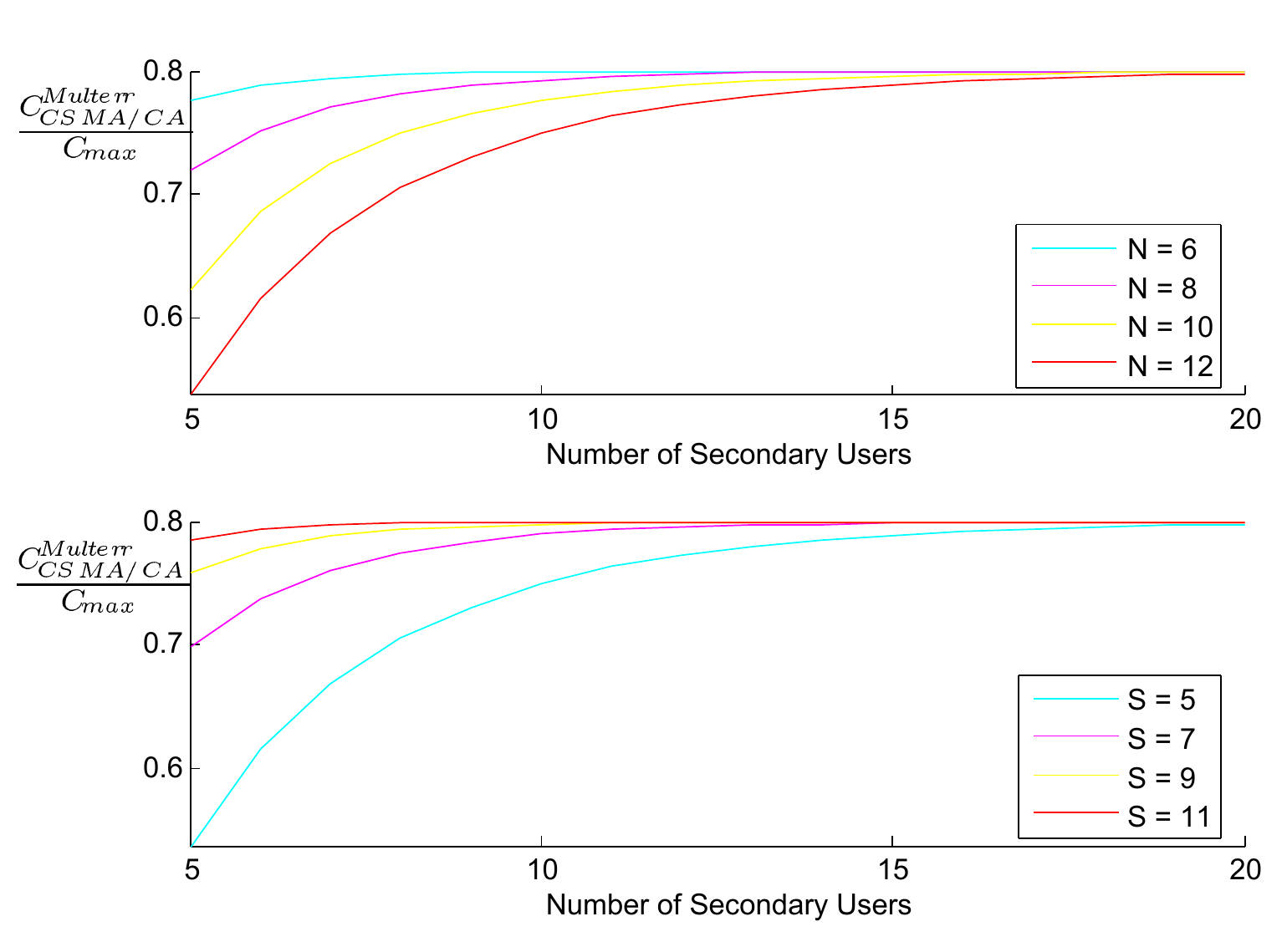}
\caption{Normalized throughput of secondary users with multiple channel erroneous sensing and accessing capability for channel utilization statistic $\rho = 0.2$ (Rarely Utilized), $\alpha = 0.2$ and $\beta = 0.8$. (a) Fixed $S = 5$ (b) Fixed $N = 12$.}
\label{fig:CMUlterrNS}
\end{figure}

Figure \ref{fig:CMUlterrNS} demonstrates the performance of the system under detection error possibility for (a) fixed $S = 5$ and various number of the frequency channels and for (b) fixed $N = 12$ and various channel sensing capability. In accordance with the previous section, as the ratio $N/S$ increases, more secondary users are needed to fully scan and utilize the left over frequency opportunities. Also we can observe that for large enough $M$, the performance of the system becomes independent of $S$.

\begin{figure}
\centering
\includegraphics[width=5.5in]{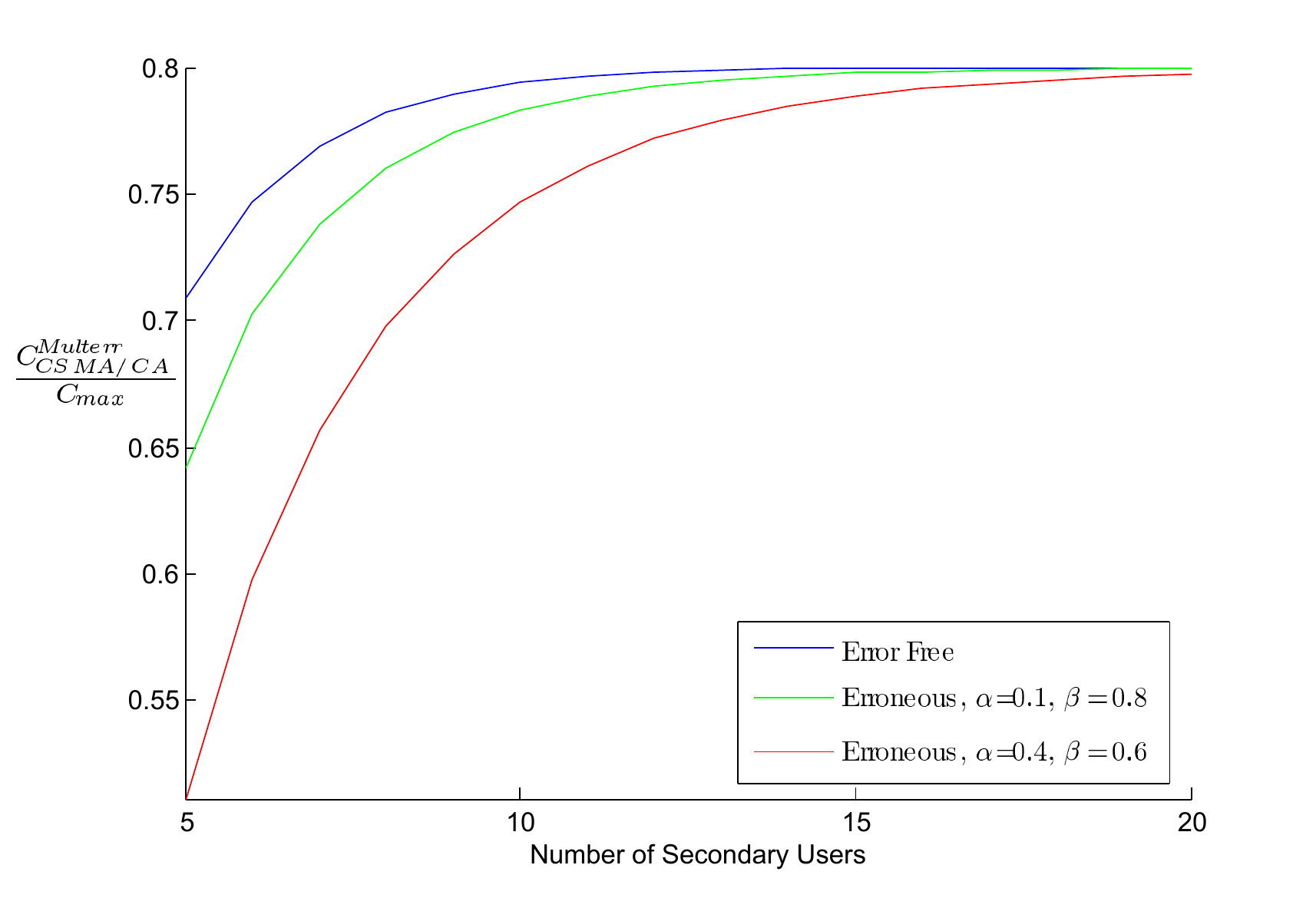}
\caption{Normalized throughput of secondary users with multiple channel sensing and accessing capability for channel utilization statistic $\rho = 0.2$ (Rarely Utilized), and erroneous detection outcome. (a) Fixed $S = 3$ (b) Fixed $N = 7$.}
\label{fig:CMUltVSerr}
\end{figure}

In Figure \ref{fig:CMUltVSerr} we compare the performance of the system for two case of error free and erroneous sensing outcomes. As shown in the figure when sensing outcome is prone to errors, secondary users have to sacrifice a portion of their throughput to guarantee the collision requirements of the primary network. As the probability of miss-detection (i.e. $\alpha$) increases and/or the probability of [correct] detection (i.e. $\beta$) decreases the decisions taken by secondary users become more unreliable which results in the decline of the secondary network throughput.

\chapter{Conclusion}
\label{conclusion}
\body

In this thesis we examined the problem of dynamic spectrum access in the presence of licensed users when the unlicensed wireless devices have intelligent radios that can adjust their transmissions in response to their environment. By detecting the underutilized bands and making use of transmission opportunities, cognitive radios show a bright future in complementing the current communications system and utilizing the frequency resources more efficiently.

In this thesis we focused on the design and performance analysis of randomized sensing and access policies. Considering the autonomous nature of the secondary users, we desire such schemes to be distributed and decentralized. In addition, we would like to avoid the use of control channels to coordinate the secondary users actions due to their implementation issues. Hence, randomized schemes are appropriate candidates to fulfill our goals.

Due to technological limitations, we made the practical assumption that the sensing capability of secondary users is limited to a subset of the available frequency bands. We addressed the problem of: How to decide on which channels or channel groups to sense out of all possible choices, to maximize the expected throughput for the ensemble of the secondary users. We demonstrated that the performance of the system improves considerably when the secondary users are equipped with wide-range detectors.

We also addressed two main factors that affect the performance of the secondary sensors and ultimately of the secondary network. First, we considered the effect of erroneous detection outcomes on the design of the sensing and access schemes in order to meet the interference specification. Second, we studied the effect of the distributed nature of the wireless ad-hoc networks on the detection outcomes. We showed that even if we manage to build perfect sensor, the detection outcomes will not be error free due to the distribution of the wireless transceivers. Thus, it does not suffice to consider the sensor ROC as the sole design factor. Finally, with detection uncertainty, we showed that secondary users can guarantee the imposed collision constraint at the cost of their aggregate throughput.

In order to resolve the contention among the secondary users we consider two MAC protocols; ALOHA vs. CSMA/CA. 
We showed that more users are always preferred in CSMA/CA-based protocols from the perspective of the total network throughput. However, an excessive number of users will deteriorate the performance of the ALOHA-based protocols. For some special cases based on ALOHA protocols, we determined the optimal number of the users that can maximize the secondary network throughput while remaining transparent to the primary network.

Although in this thesis we addressed the certain design issues for randomized cognitive MAC protocols, there are still many implementation challenges that need to be addressed before we could realize practical cognitive networks. For instance, we assumed that secondary transmitters and receivers are synchronized, i.e., the receiver is fully aware of the channel(s) that the transmitter is trying to access at any given time, which itself is a hard problem to solve. Other areas for future work are to study the performance of randomized schemes in large-scale networks and to design reliable wide-band detectors.

\end{document}